\documentclass[journal=jctcce, manuscript=article]{achemso}

\usepackage{chemformula} 
\usepackage[version=3]{mhchem}  
\usepackage[T1]{fontenc} 

\usepackage{amsmath} 
\usepackage{physics} 
\usepackage{mathrsfs} 
\usepackage{amssymb} 

\usetikzlibrary{positioning, arrows.meta} 

\usepackage{caption}
\usepackage{subcaption}

\usepackage{booktabs} 
\usepackage{threeparttable} 
\usepackage{siunitx} 
\usepackage{multirow} 

\usepackage{longtable} 

\usepackage{listings}

\author{Xiao Liu}
\affiliation{Kenneth S. Pitzer Center for Theoretical Chemistry, Department of Chemistry, University of California, Berkeley, California 94720, United States of America}
\email{xiao_liu@berkeley.edu}

\author{Kaushik D. Nanda}
\affiliation{Q-Chem, Inc., 6601 Owens Drive, Suite 240, Pleasanton, California 94588, United States of America}

\author{Jiashu Liang}
\affiliation{Kenneth S. Pitzer Center for Theoretical Chemistry, Department of Chemistry, University of California, Berkeley, California 94720, United States of America}

\author{Martin Head-Gordon}
\affiliation{Kenneth S. Pitzer Center for Theoretical Chemistry, Department of Chemistry, University of California, Berkeley, California 94720, United States of America}
\alsoaffiliation{Chemical Sciences Division, Lawrence Berkeley National Laboratory, Berkeley, California 94720, United States of America}
\email{mhg@cchem.berkeley.edu}

\title[]
  {Efficient, precise DFT calculations of NMR shieldings: Revisiting the finite field approach}


\usepackage{subfiles}

\begin{document}

\begin{tocentry}





\begin{figure}[H]
    \centering
    \includegraphics[width=0.95\linewidth]{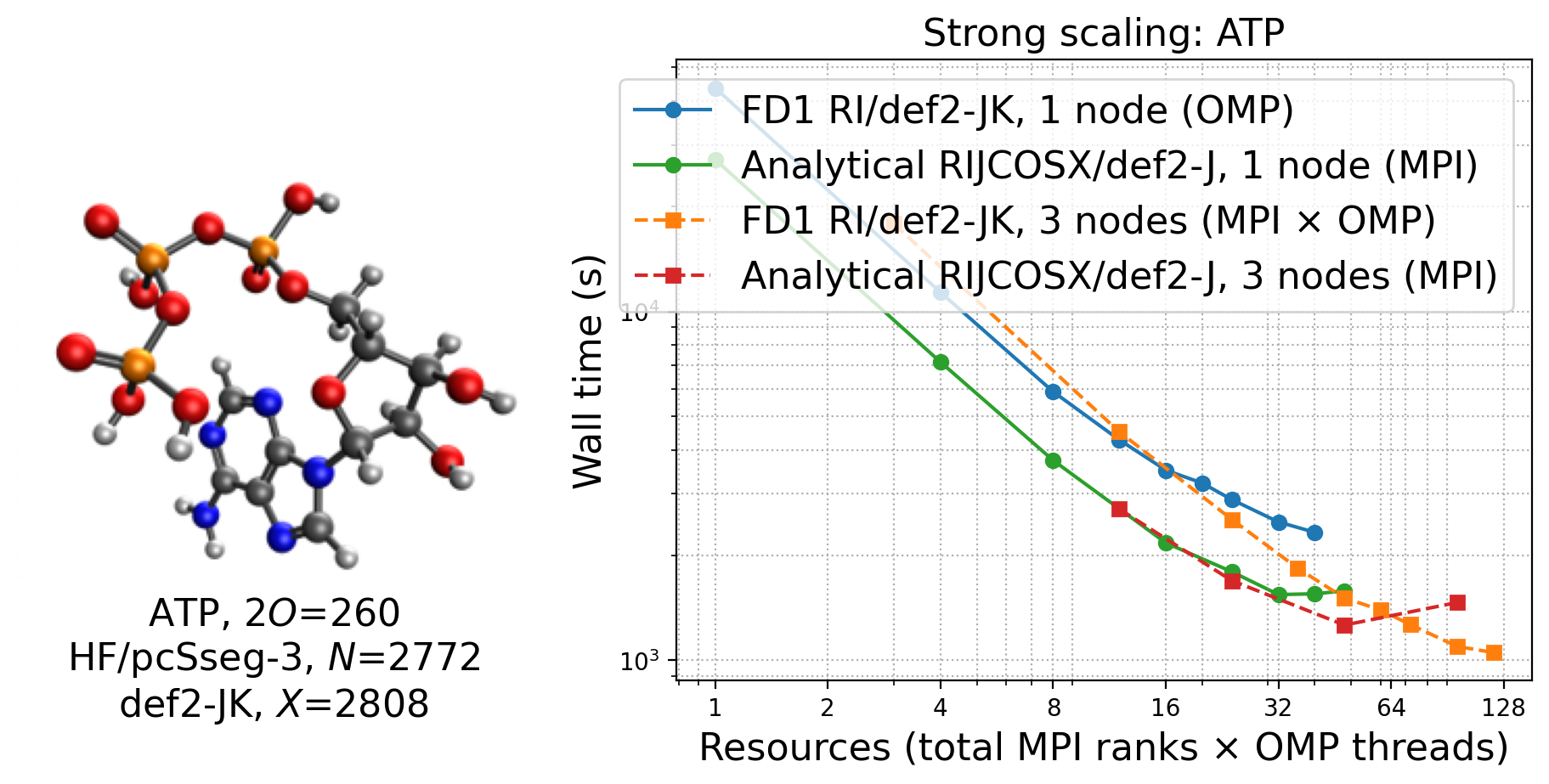}
    \label{fig:toc} 
\end{figure}

\end{tocentry}

\begin{abstract}

Absolute nuclear magnetic shielding constants and relative chemical shifts are second-derivative response properties that underpin the interpretation of Nuclear Magnetic Resonance (NMR) spectra and provide critical insights into the structural and electronic environments of diverse chemical systems. However, their accurate computation is often constrained by the need for method-specific, analytical response implementation, and therefore is particularly challenging for non-variational, correlated wavefunction methods where analytical second derivatives are frequently unavailable. Here we present a scalable framework that computes NMR shielding tensors through finite magnetic field differentiation of complex-valued, gauge-including atomic-orbital (GIAO) based self-consistent field (SCF) calculations. By combining hybrid MPI and OpenMP parallelization and resolution-of-identity (RI) approximation-based Coulomb (J) and Exchange (K) implementation and mixed numerical/analytical derivatives computational scheme, we achieve very good efficiency for hybrid density functional theory (DFT) NMR shielding, relative to existing state-of-the-art analytic implementations across realistic chemical systems of various sizes. Forward differentiation with optimal step sizes derived from rigorous error analysis retains favorable numerical errors much smaller than experimental uncertainty or intrinsic DFT errors. The results demonstrate that RI-accelerated finite magnetic field calculations can obtain DFT-level NMR shielding constants very precisely, providing a scalable foundation for extensions to more advanced quantum chemistry methods in future work.

\end{abstract}

\section{Introduction}
Nuclear magnetic resonance (NMR) spectroscopy is one of the most powerful experimental tools for molecular identification and structure elucidation. It provides a local electronic fingerprint of the molecular environment and is routinely used to study structural isomers \cite{grimblat2016computational}, stereoisomers \cite{smith2010assigning, grimblat2015beyond}, inter- and intra-molecular interactions \cite{dravcinsky2026nmr}, dynamic solvation structures \cite{you2025decoding, andrews2026dft}, crystallography \cite{harris2004nmr, paruzzo2018chemical}, and biomolecular features \cite{wuthrich2003nmr}. However, experimental NMR spectra can be difficult to interpret directly, especially for complex systems with many overlapping signals or multiple plausible structural assignments. Quantum chemistry calculations therefore play an important role in connecting candidate structures to experimentally observed signals \cite{grimblat2016computational, smith2010assigning, grimblat2015beyond, lodewyk2012computational}. 

To achieve this, one of the key properties to calculate is the nuclear magnetic shielding tensor,
\begin{equation}
    \sigma_{\alpha \beta}^K = \frac{ d^2 E\left(\mathbf{m}_{K}, \mathbf{B} \right) }{ d m_{K, \alpha} d B_{\beta} }\bigg|_{\mathbf{m}_{K}=\mathbf{0}, \mathbf{B}=\mathbf{0}} 
\label{eq:shielding}
\end{equation}
where $\mathbf{m}_{K}$ is the magnetic moment at nuclear center $K$, $\mathbf{B}$ is the external magnetic field, $\alpha$ and $\beta$ are cartesian components.
In liquid-state NMR experiments, the observed chemical shift $\delta$ is measured as an isotropic average that corresponds to the relative difference between the absolute shielding $\sigma$ of the nucleus $K$ on a target system and that on a reference molecule,
\begin{equation}
    \delta_{iso}^{tar, K} = \sigma_{iso}^{tar,K} - \sigma_{iso}^{ref, K}
\end{equation}

NMR shielding constants are most often computed \cite{lodewyk2012computational,krivdin2022computational,das2025exploring} with density functional theory (DFT), \cite{kohn1996density,capelle2006bird,perdew2009some,burke2013dft,mardirossian2017thirty} using a suitable choice of density functional approximation (DFA). Within DFT, analytical second-derivative response theory is conventionally used to evaluate the magnetic shielding tensor. This is done with the following atomic orbital (AO) density matrix-based representations \cite{ditchfield1974self, wolinski1990efficient}:
\begin{equation}
    \sigma_{\alpha \beta}^K = \sum_{\mu \nu} D_{\nu \mu} \cdot \frac{\partial^2 h_{\mu \nu}}{\partial m_{K, \alpha} \partial B_{\beta} } + \sum_{\mu \nu} \frac{\partial D_{\nu \mu}}{\partial B_{\beta} } \cdot \frac{\partial h_{\mu \nu}}{\partial m_{K, \alpha} }
\label{eq:analytical_shielding}
\end{equation}

Here $\mathbf{D}$ represents the one-particle density matrix with $\mu, \nu$ being AO indices. The first term (diamagnetic contribution) corresponds to the direct response of Hamiltonian matrix elements to both perturbations and can be calculated directly by contracting the one-particle density matrix at zero field with partial derivative integrals, whereas the second term (paramagnetic contribution) requires the density response to the external magnetic field. 

DFT is a self-consistent field (SCF) theory, and obtaining the density response requires solving coupled-perturbed SCF (CPSCF) equations\cite{ditchfield1974self,wolinski1990efficient} to obtain the magnetic field derivative of the density. For post-SCF methods, additional (non-variational) response contributions are also required and usually handled through Lagrangian and Z-vector formalisms \cite{handy1984evaluation}. Dynamic correlation treatments using M\o ller-Plesset perturbation theory (MPPT) \cite{gauss1993effects, gauss1992calculation, gauss1994giao}, Random Phase Approximation (RPA) \cite{drontschenko2023analytical}, Coupled Cluster (CC) theory up to the inclusion of full triples \cite{gauss1996perturbative, gauss2000analytic, gauss2002analytic}, as well static correlation treatments using active space-based methods \cite{ruud1994multiconfigurational, nottoli2022computation} or strong correlation corrected local hybrid functionals \cite{schattenberg2024implementation} have been achieved with analytical second-derivative response theory-based NMR shielding calculations. Developments with approximated two-electron integrals for efficient Coulomb (J) and/or Exchange (K) build \cite{cao2005nuclear, loibl2010density, stoychev2018self} or the correlation contribution \cite{stoychev2018efficient, burger2021nmr, nottoli2022computation}, non-self-consistent density response (in a ``density-corrected'' DFT manner) \cite{lai2026search}, low scaling methods with AO-driven algorithms or local correlation approaches \cite{kussmann2007linear, beer2011nuclei, maurer2013linear, gauss2000nmr, loibl2012nmr, stoychev2021dlpno, glasbrenner2021efficient} have also been achieved for NMR shielding.

The analytical second-derivative approaches are rigorous and efficient when available, but their implementation is algebraically demanding, method-specific, and often difficult to parallelize efficiently, especially in distributed-memory MPI environments such as modern massively parallel supercomputer architectures. A fully numerical finite-difference strategy provides the opposite tradeoff. One may displace both the nuclear magnetic moment and the external magnetic field and approximate the shielding tensor directly from total energy differences,
\begin{equation}
    \sigma_{\alpha \beta}^K \approx \frac{ E(\Delta m_{K, \alpha}, \Delta B_{\beta}) - E(\Delta m_{K, \alpha}, -\Delta B_{\beta}) - E(-\Delta m_{K, \alpha}, \Delta B_{\beta}) + E(-\Delta m_{K, \alpha}, -\Delta B_{\beta}) }{4 \Delta m_{K, \alpha} \Delta B_{\beta}}
\end{equation}

Such a finite-difference second-derivative scheme is straightforward and sometimes the only practical option when analytical first derivatives are also unavailable. Following the original work by Fukui et al. in the 1990s \cite{fukui1992calculation, fukui1994calculation}, the finite magnetic field approach has been recently applied to study novel wavefunction methods such as $\kappa$-regularized second-order M\o ller-Plesset perturbation theory ($\kappa$-MP2) \cite{wong2023silico}, direct Random Phase Approximation (d-RPA) and $\sigma$-functionals \cite{glasbrenner2021benchmarking, fauser2024accurate}, alongside other novel applications on molecular properties or dynamics in static magnetic fields \cite{hampe2017equation, hampe2019transition, blaschke2022cholesky, tellgren2009non, lange2012paramagnetic, sen2019excited, monzel2022molecular, bhati2025phase, sun2019ab, tang2024exact, tang2025simulating, tang2026ab}. However, this approach is computationally inefficient for full system shielding constant calculations: evaluating isotropic shieldings for all nuclei requires nuclear magnetic moment displacements for each atom, leading to a cost proportional to $3 \times 4 \times n_{atoms}$. In addition, the numerical accuracy depends simultaneously on two finite-difference step sizes of the finite magnetic field and finite nuclear magnetic moment, and could lead to large numerical errors if the step sizes are not chosen properly.

In this work, we develop a mixed analytical and numerical strategy that combines the main advantages of fully analytical and numerical approaches while mitigating their limitations. This is an old idea, first discussed for polarizabilities\cite{cohen1965electric} (and subsequently very widely applied: see e.g. ref. \citenum{hait2018accurate}). The mixed strategy of one analytical and one numerical derivative was generalized by Pople et al.\cite{pople1968self} who also applied it to magnetic shielding.\cite{ditchfield1970molecular} By reformulating the second derivative property evaluation as a finite first difference of a first derivative property, we gain several advantages. First, we require only analytic first derivatives, and thus gain simplicity and easier generalization to higher level methods  (although there is a non-trivial effort to instrument quantum chemistry software to perform calculations in finite magnetic fields\cite{tellgren2008nonperturbative, reynolds2015fully, stopkowicz2015coupled, wong2023silico}). Second, the embarrassingly parallel nature of numerical differentiation allows us to design an efficient hybrid parallel algorithm with distributed memory (MPI) for each field and shared memory (OpenMP) for a given field strength. 

Our objective is to explore how viable the mixed numerical and analytical strategy can be for DFT-level calculations of NMR shieldings relative to the fully analytical approach. We take advantage of efficient resolution-of-identity (RI) approximation-based algorithms that have been successfully applied to SCF energies and first derivatives, such as RI-JK \cite{weigend2002fully} and occ-RI-K \cite{manzer2015fast}.  We discuss the theory, with careful focus on error analysis, followed by details of the parallel algorithm, and our present implementation of RI-accelerated complex SCF in finite magnetic fields. This implementation, together with our parallel algorithm, achieves very good computational efficiency on realistic chemical systems of different sizes. Finally, we demonstrate the utility of our efficient algorithm on a natural product, hexacyclinol, whose structure was historically the subject of significant controversy. 

\section{Theory and algorithm design}
\subsection{Molecular Hamiltonian in finite magnetic field and GIAO basis}
\label{sec:hamiltonian}

Here we present a brief overview of the canonical theory for NMR shielding calculations. For more thorough discussions, we refer the readers to the extensive review by Helgaker et al. \cite{helgaker1999ab} and the more recent review by Das and Merz \cite{das2025exploring}. The full non-relativistic, closed-shell Hamiltonian for NMR shielding calculation up to first order in $\mathbf{m}_{K}$ and first order in $\mathbf{B}$, which we shall denote as $\hat{H}( \mathbf{m}_{K}^{(1)}, \mathbf{B}^{(1)})$, in atomic units is \cite{ditchfield1974self, helgaker1991electronic},
\begin{equation}
\begin{aligned}
    \hat{H}( \mathbf{m}_{K}^{(1)}, \mathbf{B}^{(1)}) = & \ -\frac{1}{2} \sum_{i} \nabla^2_i - \sum_{i} \sum_{K} \frac{Z_K}{r_{iK}} + \sum_{ij} \frac{1}{r_{ij}} + \frac{1}{2} \sum_{i} \mathbf{B} \cdot \hat{\mathbf{L}}_{i} + \alpha^2 \sum_{i} \sum_{K} \frac{\mathbf{m}_{K} \cdot \hat{\mathbf{L}}_{iK}}{r_{iK}^3} \\
    & + \frac{\alpha^2}{2} \sum_{i} \sum_{K} \frac{ (\mathbf{B} \cdot \mathbf{m}_{K}) (\mathbf{r}_{i} \cdot \mathbf{r}_{iK}) - (\mathbf{B} \cdot \mathbf{r}_{iK})(\mathbf{r}_{i} \cdot \mathbf{m}_{K}) } {r_{iK}^3} \\
\end{aligned}
\end{equation}
Here $\alpha$ is the fine structure constant and $\hat{\mathbf{L}}$ is the angular momentum operator. As foreshadowed above, in the conventional, fully analytical implementation, the derivatives with respect to the nuclear magnetic moment and external magnetic field are calculated perturbatively after solving the field-free SCF. To remove the gauge origin dependence in the final observable, gauge-invariant (or ``gauge-including'') atomic orbitals (GIAOs) are often used for the response equations \cite{london1937theorie, ditchfield1974self, helgaker1991electronic}. For an ordinary, real-valued Gaussian atomic orbital $\chi_\mu$ centered on atom $K$, the corresponding GIAO basis function is,
\begin{equation}
    \omega_\mu(\mathbf{r}; \mathbf{B}, \mathbf{R}_K) = \exp(-i \mathbf{A}_{K} \cdot \mathbf{r}) \cdot \chi_\mu (\mathbf{r}; \mathbf{R}_K) = \exp[ -\frac{i}{2}\mathbf{B} \times \left( \mathbf{R}_{K} - \mathbf{R}_{O} \right) \cdot \mathbf{r}] \cdot \chi_\mu (\mathbf{r}; \mathbf{R}_K)
\end{equation}
where $\mathbf{A}_N$ corresponds to the local vector potential at the nucleus $K$ from the external magnetic field $\mathbf{B}$. At a finite magnetic field, the exponent acts like a position-dependent complex-valued phase factor and the GIAOs are therefore also complex-valued. The gauge origin dependence in the vector potential cancels out when evaluating the integrals over GIAO pairs, making the final energy and property gauge independent. 

\subsection{Finite first derivative-based shielding and its error analysis}
The Hamiltonian above omits the electronic spin Zeeman interaction, which is primarily relevant for paramagnetic NMR measurements on open-shell systems \cite{novotny2024paramagnetic}. For closed-shell systems, the net electronic spin Zeeman interaction is zero. Therefore, the $\hat{H}(\mathbf{B}^{(1)})$ interaction with the external magnetic field in the Hamiltonian, i.e., the orbital Zeeman interaction, is purely imaginary. For non-relativistic, closed-shell NMR shielding calculation with analytical second-derivative response theory (Eq. \ref{eq:analytical_shielding}), the imaginary unit in the response equation is typically factored out and the working equation can be formulated in a real-valued, anti-symmetric manner to avoid complex algebra.  

Alternatively, we can include magnetic interactions up to first order in $\mathbf{B}$, denoted as $\hat{H}(\mathbf{B}^{(1)})$, variationally in the Hamiltonian, and then calculate the first order nuclear magnetic moment response perturbatively (i.e. analytically). For a formal error analysis with $\hat{H}(\mathbf{B}^{(1)})$, we treat the external magnetic field as a finite first-order perturbation and the rest as the zeroth-order Hamiltonian $\hat{H}_{0}$,
\begin{equation}
\begin{aligned}
    \hat{H}(\mathbf{B}^{(1)}) = & -\frac{1}{2} \sum_{i} \nabla^2_i - \sum_{i} \sum_{K} \frac{Z_K}{r_{iK}} + \sum_{ij} \frac{1}{r_{ij}} + \frac{1}{2} \sum_{i} \mathbf{B} \cdot \hat{\mathbf{L}}_{i} \\
    = & \hat{H}_{0} + \mathbf{B} \cdot \hat{\mathbf{H}}^{(1)} 
\end{aligned}
\end{equation}
The zeroth-order solution, $\Psi^{(0)}$, is just the real-valued wavefunction at zero field, while the first-order wavefunction response, $\Psi^{(1)}$, is caused by the purely imaginary $\hat{\mathbf{H}}^{(1)}$ and also purely imaginary:
\begin{equation}
    \hat{\mathbf{H}}^{(1)} = \frac{1}{2} \sum_{i} \hat{\mathbf{L}}_{i} = -\frac{i}{2} \sum_{i} \mathbf{r}_{i} \times \boldsymbol{\nabla}_{i}
\end{equation}
The full Hamiltonian $\hat{H}(\mathbf{B}^{(1)})$ is now complex and requires complex SCF with the GIAO basis to enforce gauge independence. The Roothaan equation also involves complex algebra (note that the elements of $\boldsymbol{\epsilon}(\mathbf{B})$ are still strictly real due to Hermitian symmetry) and takes the fully magnetic field-dependent form,
\begin{equation}
    \mathbf{F}(\mathbf{B}) \mathbf{C}(\mathbf{B})= \mathbf{S}(\mathbf{B}) C(\mathbf{B}) \boldsymbol{\epsilon}(\mathbf{B})
\end{equation}

From expanding the solution in terms of $\mathbf{B}$, we also see that the first-order contribution to the energy is zero, corresponding to the physical fact that the expectation value of the orbital angular momentum is zero for a closed-shell system. This is due to time-reversal symmetry, and all the odd-order contributions with respect to B are zero for the same reason.
\begin{equation}
\begin{aligned}
    E(\mathbf{B}) & = E_{0} + \mathbf{B} \bra{\Psi^{(0)}} \hat{\mathbf{H}}^{(1)}  \ket{\Psi^{(0)}} + \sum_{\alpha, \beta \in (x,y,z)} B_{\alpha} B_{\beta} \sum_{I \neq 0} \frac{ \bra{ \Psi^{(0)} } \hat{H}_{\alpha}^{(1)}  \ket{ \Psi_{I}^{(0)}} \bra{\Psi_{I}^{(0)}} \hat{H}_{\beta}^{(1)} \ket{ \Psi^{(0)} } }{E^{(0)} - E_{I}^{(0)}} + O(\mathbf{B}^3)\\
    & = E_{0} + \sum_{\alpha, \beta \in (x,y,z)} B_{\alpha} B_{\beta} \sum_{I \neq 0} \frac{ \bra{ \Psi^{(0)} } \hat{H}_{\alpha}^{(1)}  \ket{ \Psi_{I}^{(0)}} \bra{\Psi_{I}^{(0)}} \hat{H}_{\beta}^{(1)} \ket{ \Psi^{(0)} } }{E^{(0)} - E_{I}^{(0)}} + O(\mathbf{B}^4)
\end{aligned}
\end{equation}
The vanishing of the odd-order terms has important implications for our finite-field algorithm and will be revisited later in this section.

The key idea of this work is to replace analytical CPSCF for magnetic response by a finite difference scheme over an analytical first derivative property, which can be obtained from finite-field GIAO-SCF without solving the perturbative response equations. Herein, the atom-wise induced magnetic field $B_\text{ind}\left( \mathbf{B} \right)$ at finite external magnetic field is calculated analytically as the first derivative of the energy with respect to a nuclear magnetic moment. With a variationally converged field-dependent GIAO-SCF solution, it can be calculated as a density-partial derivative integral contraction in AO representation,
\begin{equation}
\begin{aligned}
    B_{\text{ind}, \alpha}^K \left( \mathbf{B} \right) = & - \frac{ d E\left(\mathbf{m}_{K}, \mathbf{B} \right) }{d m_{K, \alpha}} \bigg|_{\mathbf{m}_{K}=\mathbf{0}} \\
    = & - \sum_{\mu \nu} D_{\nu \mu} \left( \mathbf{B} \right) \cdot \frac{\partial h_{\mu \nu} \left( \mathbf{B} \right)}{\partial m_{K, \alpha} }
\end{aligned}
\label{eq:FD1_b_ind}
\end{equation}
Note that the field-dependent density matrix $\mathbf{D} \left( \mathbf{B} \right) = \mathbf{C}_\text{occ}\left( \mathbf{B} \right) \mathbf{C}^\dagger_\text{occ}\left( \mathbf{B} \right)$ and the GIAO partial derivative integrals with respect to nuclear magnetic moment are both complex-valued. The working equations to evaluate the GIAO partial derivative integrals can be found in the appendix. The shielding tensor, Eq. \ref{eq:shielding}, can then be calculated from numerical differentiation (central or forward difference) of the first derivative property, $B_{\text{ind}, \alpha}^K \left( \mathbf{B} \right)$ with respect to the finite external magnetic field strength $\mathbf{B}$. This first derivative finite difference procedure will be referred to as ``FD1''. The forward and central difference expressions are standard (recalling that $\mathbf{B}_{\text{ind}}^K \left( \mathbf{B}=\mathbf{0} \right) = \mathbf{0}$):
\begin{align}
    \sigma_{\alpha \beta}^{K, \text{FD1-forward}} &= - \frac{B^{K}_{ind, \alpha}(\Delta B_{\beta})}{\Delta B_{\beta}} \\
    \sigma_{\alpha \beta}^{K, \text{FD1-central}} &= - \frac{B^{K}_{ind, \alpha}(\Delta B_{\beta}) - B^{K}_{ind, \alpha}(-\Delta B_{\beta})}{2\Delta B_{\beta}}
\end{align}
For a general expansion over an energy derivative property, the forward difference formally contains a first-order truncation error, whereas the leading truncation error from the central difference is second order. Naively, this would suggest that the central difference approach is required for high accuracy. However, for closed-shell systems with time-reversal symmetry, the total energy obeys,
\begin{equation}
    E(+\Delta \mathbf{m}_{K}, +\Delta \mathbf{B}) = E(-\Delta \mathbf{m}_{K}, -\Delta \mathbf{B})
\end{equation}
Equivalently, if the energy is expanded as,
\begin{equation}
    E(\mathbf{m}_{K}, \mathbf{B}) = \sum_{mn} \mathbf{m}_{K}^{m} : \mathbf{B}^{n} : \frac{ \partial^{m+n} E\left(\mathbf{m}_{K}, \mathbf{B} \right) }{ (\partial \mathbf{m}_{K})^{m} (\partial \mathbf{B})^{n} }\bigg|_{\mathbf{m}_{K}=\mathbf{0}, \mathbf{B}=\mathbf{0}}
\end{equation}
then only the terms with even $m+n$ powers would survive. As a result, the leading truncation error of the forward difference first derivative-based shielding is also second order in $\Delta B_{\beta}$,
\begin{equation}
\begin{aligned}
    \sigma_{\alpha \beta}^{K}
    = & \sigma_{\alpha \beta}^{K, \text{FD1-forward}} + \sum_{\gamma \in (x,y,z)} \frac{\Delta B_{\gamma}}{2} \frac{\partial^3 E}{\partial m_{K, \alpha} \partial B_{\beta} \partial B_{\gamma}} + \sum_{\gamma,\delta \in (x,y,z)} \frac{ \Delta B_{\gamma} \Delta B_{\delta} }{6} \frac{\partial^4 E}{\partial m_{K, \alpha} \partial B_{\beta} \partial B_{\gamma} \partial B_{\delta}} \\ 
    & + \sum_{\gamma,\delta,\epsilon \in (x,y,z)}  O( \Delta B_{\gamma} \Delta B_{\delta} \Delta B_{\epsilon} ) \\
    = & \sigma_{\alpha \beta}^{K, \text{FD1-forward}} +\sum_{\gamma,\delta \in (x,y,z)} \frac{ \Delta B_{\gamma} \Delta B_{\delta} }{6} \frac{\partial^4 E}{\partial m_{K, \alpha} \partial B_{\beta} \partial B_{\gamma} \partial B_{\delta}} + \sum_{\gamma,\delta,\epsilon,\zeta \in (x,y,z)} O( \Delta B_{\gamma} \Delta B_{\delta} \Delta B_{\epsilon} \Delta B_{\zeta} ) \\
\end{aligned}
\end{equation}

At a practical level this means that only half as many finite $\mathbf{B}$ field calculations (3) can be performed via forward differences as for central differences (6), with no formal consequence for the resulting finite difference error.

The symmetry basis of this conclusion is known: for non-degenerate time-reversal-symmetric systems, magnetic-response expansions contain only the symmetry-allowed powers \cite{pennanen2008nuclear, van2012nmr}, and field-dependent closed-shell NMR shielding corrections first predicted by Ramsey in 1970 \cite{ramsey1970possibility} indeed begin at $B^2$, according to a recent experimental confirmation \cite{kantola2020direct}. However, to our knowledge, the practical consequence that a one-sided finite-field differentiation can exhibit the same leading order truncation error as the corresponding central difference has not been explicitly emphasized in prior literature. In a similar vein, this conclusion could also be extended to the calculations of other magnetic properties such as the magnetizability, J coupling constants, the rotational g tensor, and nuclear spin-rotation constants, so long as the system of interest is genuinely closed-shell and the unperturbed zeroth-order Hamiltonian is fully time-reversal symmetric.

\subsection{Hybrid MPI/OpenMP parallel algorithm design}
We have established that the forward difference scheme could reduce the number of expensive complex-valued GIAO-SCF calculations by a factor of two (from 6 to 3) compared to central differences while preserving the same numerical accuracy. How should those expensive calculations be performed?  The first issue is the initial guess. At the very weak fields used for the finite differences ($\sim 10^{-3}-10^{-4}$ a.u. as discussed later) the optimized real-valued GTO MO coefficients at zero magnetic field (i.e., $\mathbf{C}(\mathbf{B}=0)$) provide a very good initial guess for $\mathbf{C}(\mathbf{B})$ to accelerate complex SCF with GIAOs. This ``orbital recycling'' logic can give substantial reductions in the number of SCF iterations needed for all 3 finite field calculations. Therefore we first converge the field-free SCF.

We can then take advantage of natural parallel structure in the finite-field first-derivative formulation. Each of the 3 finite magnetic field calculations and $ B^{K}_{\text{ind}, \alpha}(\Delta B_{\beta}) $ evaluations is independent and no communication is required until the final assembly of the shielding tensor, which is almost ideal for MPI parallelization. The orbital recycling logic also helps with load balancing for normal finite molecular systems with little anisotropy in finite magnetic field, in the sense that the GIAO-SCF for $\mathbf{B} = (\Delta B, 0, 0)$, or $(0, \Delta B, 0)$, or $(0, 0, \Delta B)$ takes almost the same number of SCF iterations and thus almost the same wall time on each MPI rank, with each one starting from the same high quality zero field initial guess. 

We therefore distribute the 3 independent finite magnetic field calculations on 3 MPI ranks. On each rank, OpenMP is applied to accelerate complex-valued GIAO-SCF and the contraction of the density matrix and partial derivative integral (Eq. \ref{eq:FD1_b_ind}). Finally, after the atom-wise induced magnetic field calculations are ready on the 3 independent MPI ranks, we do one MPI gather step to calculate the shielding for the full system. As the induced magnetic field $\mathbf{B}_\text{ind}$ is only a vector of length $3*n_\text{atoms}$, this algorithm has a very small MPI communication cost that scales $\mathcal{O}(n_\text{atoms})$ with system size versus the roughly $\mathcal{O}(n_\text{atoms}^2)-(n_\text{atoms}^4)$ scaling of the SCF calculations.

The full algorithm is illustrated below in Figure \ref{fig:hybrid-mpi-openmp}. Note that in this work, we only focus on an efficient implementation for density functionals up to rung 4 (including Hartree-Fock as a special case). The same algorithm structure could be extended to correlated post-SCF methods if they are appropriately instrumented for finite magnetic fields, with only a requirement for analytic first derivatives.

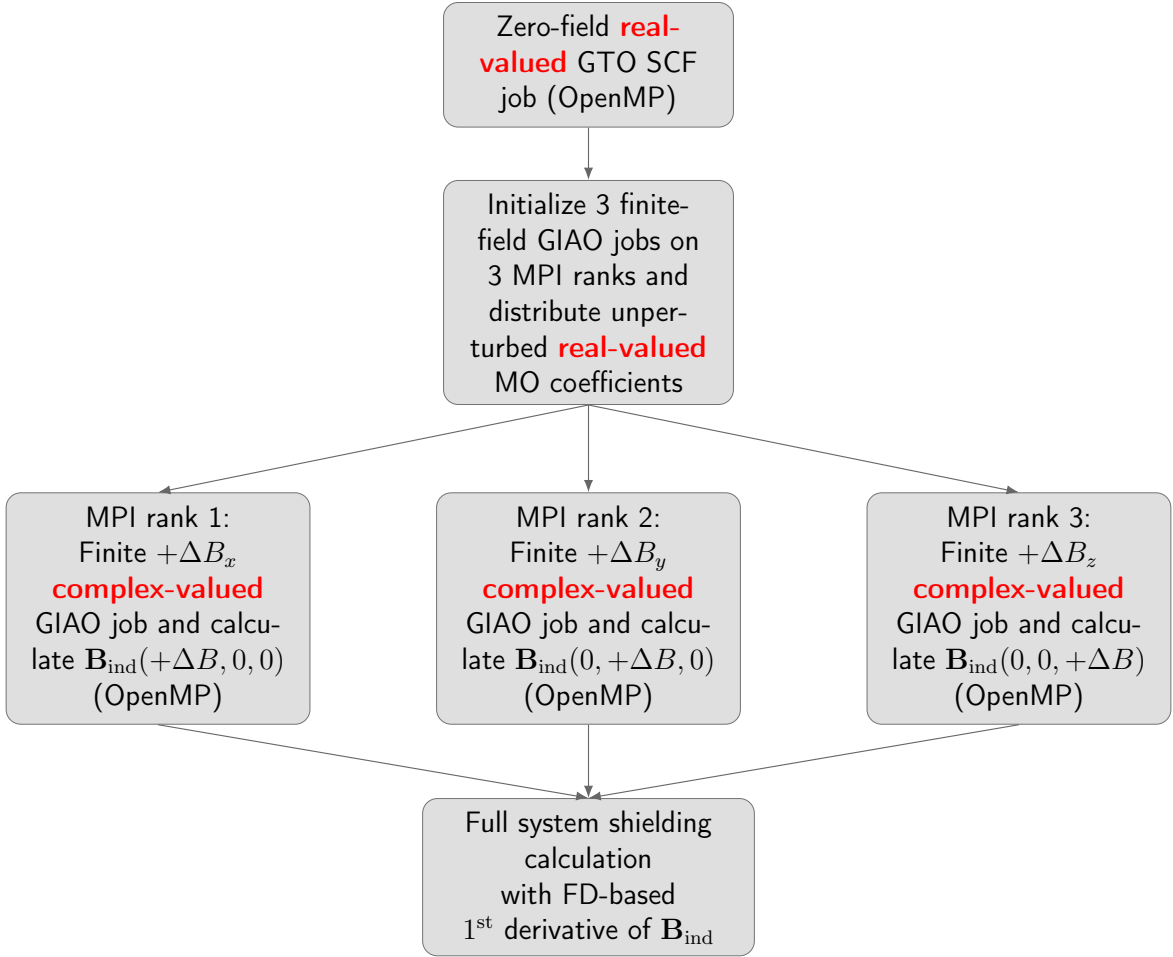
\begin{figure}[H]
\centering
\resizebox{0.95\linewidth}{!}{
\begin{tikzpicture}[
    font=\sffamily,
    >=Latex,
    box/.style={
        draw=black!55,
        rounded corners=6pt,
        fill=gray!25,
        align=center,
        text width=3.8cm,
        minimum height=1.1cm,
        inner sep=5pt
    },
    rankbox/.style={
        box,
        text width=4.0cm,
        minimum height=1.25cm
    },
    arrow/.style={
        ->,
        draw=black!60,
        line width=0.45pt
    }
]

\node[
    rectangle,
    minimum width=15.5cm,
    minimum height=8.5cm,
    anchor=north
] at (0,0.2) {};

\node[box] (gto) at (0,0)
{Zero-field \textcolor{red}{\textbf{real-valued}} GTO SCF job (OpenMP)};

\node[box, below=0.75cm of gto] (mo)
{Initialize 3 finite-field GIAO jobs on 3 MPI ranks and distribute unperturbed \textcolor{red}{\textbf{real-valued}} MO coefficients};

\node[rankbox, below left=1.25cm and 1.9cm of mo] (r1)
{MPI rank 1:\\
Finite $+\Delta B_x$ \textcolor{red}{\textbf{complex-valued}} GIAO job and calculate $\mathbf{B}_{\mathrm{ind}}(+\Delta B, 0, 0)$ (OpenMP)};

\node[rankbox, below=1.25cm of mo] (r2)
{MPI rank 2:\\
Finite $+\Delta B_y$ \textcolor{red}{\textbf{complex-valued}} GIAO job and calculate $\mathbf{B}_{\mathrm{ind}}(0, +\Delta B, 0)$ (OpenMP)};

\node[rankbox, below right=1.25cm and 1.9cm of mo] (r3)
{MPI rank 3:\\
Finite $+\Delta B_z$ \textcolor{red}{\textbf{complex-valued}} GIAO job and calculate $\mathbf{B}_{\mathrm{ind}}(0, 0, +\Delta B)$ (OpenMP)};

\node[
    box,
    text width=4.4cm,
    below=1.05cm of r2
] (final)
{Full system shielding\\
calculation with FD-based\\
$1^{\mathrm{st}}$ derivative of $\mathbf{B}_{\mathrm{ind}}$
};

\draw[arrow] (gto) -- (mo);

\draw[arrow] (mo.south) -- (r1.north);
\draw[arrow] (mo.south) -- (r2.north);
\draw[arrow] (mo.south) -- (r3.north);

\draw[arrow] (r1.south) -- (final.north);
\draw[arrow] (r2.south) -- (final.north);
\draw[arrow] (r3.south) -- (final.north);

\end{tikzpicture}
}
\caption{Hybrid MPI/OpenMP parallel algorithm for finite-field shielding calculations.}
\label{fig:hybrid-mpi-openmp}
\end{figure}

\subsection{RI-JK and occ-RI-K with complex-valued GIAO primary basis}
\label{sec:giao-ri}
We next need to instrument an efficient Fock build algorithm for evaluation in finite magnetic fields, in the GIAO basis. We elected to employ the resolution-of-identity (RI) approximation for the expensive complex-valued GIAO-SCF to accelerate the Coulomb (J) and Exchange (K) matrix assembly. As established by Reynolds and Shiozaki \cite{reynolds2015fully}, and Pausch and Klopper \cite{pausch2020efficient}, for complex-valued GIAO calculations in a finite magnetic field, using real-valued GTO auxiliary basis can restore the same 4-fold permutational symmetry in the mixed 2-electron, 3-center (2e3c) integrals as the fully real-valued analog. Working in the Coulomb metric\cite{vahtras1993integral}, and using indexes $P,Q$ to represent auxiliary basis functions ($X$ in total):
\begin{equation}
    ( \omega_{\mu} \omega_{\nu} | \omega_{\lambda} \omega_{\sigma} ) \approx \sum_{PQ}^X ( \omega_{\mu} \omega_{\nu} | \chi_{P} ) ( \chi_{P} | \chi_{Q} )^{-1} ( \chi_{Q} | \omega_{\lambda} \omega_{\sigma} )
\end{equation}
The 4-fold permutational symmetry in the mixed 2e3c integrals listed below is the same as the real-valued GTO analog (asterisk superscript denotes complex conjugate),
\begin{equation}
    ( \omega_{\mu} \omega_{\nu} | \chi_{P} )
    = ( \omega_{\nu} \omega_{\mu} | \chi_{P} )^{*}
    = ( \chi_{P} | \omega_{\mu} \omega_{\nu} )
    = ( \chi_{P} | \omega_{\nu} \omega_{\mu} )^{*}
\end{equation}

Here we use the same strategy to implement the RI-JK algorithm\cite{weigend2002fully} and occ-RI-K algorithm\cite{manzer2015fast} with complex-valued GIAO primary basis and real-valued GTO auxiliary basis. The underlying mixed GIAO-GTO 2e3c integrals are evaluated with the Head–Gordon–Pople (HGP88) algorithm \cite{head1988method}. RI-K and occ-RI-K algorithms with the range-separated hybrid exchange operator as needed for range-separated hybrid functionals\cite{mardirossian2017thirty} are also implemented, in addition to the conventional exact (Hartree-Fock) exchange for global hybrids. For RI-J and RI-K, the GIAO adaptation is relatively straightforward,

\begin{equation}
\begin{aligned}
    J_{\mu \nu}
    = & \sum_{\lambda \sigma}^N ( \omega_{\mu} \omega_{\nu} | \omega_{\sigma} \omega_{\lambda} ) P_{\lambda \sigma} \\
    \approx & \sum_{\lambda \sigma}^N \sum_{PQ}^X ( \omega_{\mu} \omega_{\nu} | \chi_{P} ) ( \chi_{P} | \chi_{Q} )^{-1} ( \chi_{Q} | \omega_{\sigma} \omega_{\lambda} ) P_{\lambda \sigma} \\
\end{aligned}    
\end{equation}

\begin{equation}
\begin{aligned}
    K_{\mu \nu}
    = & \sum_{\sigma \lambda}^N ( \omega_{\mu}  \omega_{\lambda} |  \omega_{\sigma}  \omega_{\nu} ) P_{\lambda \sigma} \\
    \approx & \sum_{\sigma \lambda}^N \sum_{PQ}^X (  \omega_{\mu}  \omega_{\lambda} | \chi_{P} ) ( \chi_{P} | \chi_{Q} )^{-1} ( \chi_{Q} |  \omega_{\sigma}  \omega_{\nu} ) P_{\lambda \sigma} \\
\end{aligned}    
\end{equation}

For occ-RI-K, the computational saving over the conventional RI-K comes from projection on the exchange operator and ignoring the virtual-virtual component in the projection expansion. As established in the original occ-RI-K literature \cite{manzer2015fast}, the inexact virtual-virtual block in the exchange matrix does not affect the SCF energy (which only requires the exact occupied-occupied block) or first derivative properties such as the nuclear gradient (only requires exact occupied-virtual block) as well as the error vector used in direct inversion of the iterative subspace (DIIS) \cite{pulay1980convergence, pulay1982improved}. However, as a feature of this algorithm, the inexact virtual-virtual matrix elements in the Fock matrix does affect elements of the orbital hessian. Analytical computation of second derivative properties is hence affected through the hessian-vector product form to converge the Z vector during the CPSCF linear solve step, where occ-RI-K is thus not applicable. 

Here we circumvent that issue by design: the analytical first derivative property $\mathbf{B}_\text{ind}\left( \mathbf{B} \right)$ does not depend on the orbital hessian, and using first finite differences to calculate the second derivative property avoids constructing the orbital hessian-vector product. Therefore, the projection treatment in occ-RI-K does not introduce new errors into the final FD1 NMR shielding and is just as accurate as RI-K itself. 
Of course the same first finite difference strategy can be applied to other second derivative properties such as the vibrational frequency or polarizability with the conventional occ-RI-K algorithm. 

The working equation for the GIAO-adapted occ-RI-K algorithm can be expressed as follows: 
\begin{equation}
    K^\text{occ-RI-K}_{\mu \nu} =
    \sum_{i}^{O} \sum_{\lambda}^N S_{\mu \lambda} c_{\lambda i} K_{i \nu} +
    \sum_{i}^{O} \sum_{\lambda}^N K_{\mu i} c_{\lambda i}^{*} S_{\lambda \nu} -
    \sum_{i, j}^{O} \sum_{\lambda, \sigma}^N S_{\mu \lambda} c_{\lambda i} K_{i j} c_{\sigma j}^{*} S_{\sigma \nu}
\end{equation}
Here $O$ is the number of occupied orbitals, and $K_{i \nu}$ are elements of the mixed occupied MO-AO exchange matrix,
\begin{equation}
\begin{aligned}
    K_{i \nu} = & \sum_{j}^{O} (\omega_{i} \omega_{j} | \omega_{j} \omega_{\nu}) \\
    = & \sum_{j}^{O} \sum_{\mu, \lambda, \sigma}^N c_{\mu i}^{*} c_{\lambda j} c_{\sigma j}^{*} (\omega_{\mu} \omega_{\lambda} | \omega_{\sigma} \omega_{\nu}) \\
    \approx & \sum_{j}^{O} \sum_{\mu, \lambda, \sigma}^N  c_{\mu i}^{*} c_{\lambda j} c_{\sigma j}^{*} (\omega_{\mu} \omega_{\lambda} | \chi_{P}) (\chi_{P}|\chi_{Q})^{-1}  (\chi_{Q} | \omega_{\sigma} \omega_{\nu})
\end{aligned}
\end{equation}
The occupied-occupied exchange matrix elements $K_{i j}$ are simply obtained from an AO to MO transformation of the second index:
\begin{equation}
    K_{i j} = \sum_{\nu}^N K_{i \nu} c_{\nu j}
\end{equation}

It is worth noting that the speedup ratio of the rate-determining $O(N^4)$ operations between conventional RI-K and occ-RI-K is,
\begin{equation}
    \frac{ N_{ \text{RI-K} } }{N_{ \text{occ-RI-K} }} = \frac{ N (N + X) }{ O (N + 2X) }
\end{equation}
where $O, N, X$ represent the number of occupied orbitals, primary basis functions and auxiliary basis functions respectively. For typical auxiliary basis sets used in a normal general purpose calculation, the speedup corresponds to a conservative estimation of $ X \approx 2 N $ is $\frac{3 N}{4 O}$. For NMR shielding calculations, the corresponding special purpose basis set (such as the Jensen pcS \cite{jensen2008basis} and pcSseg \cite{jensen2015segmented} family) are designed to provide greater radial flexibility in the low-angular-momentum parts (espcially the p functions) than typical general purpose basis sets (such as the Karlsruhe basis sets \cite{weigend2005balanced} and the Dunning basis sets \cite{dunning1989gaussian, gauss1993effects}) in comparable $\zeta$ quality. These are often used with relatively small auxiliary basis sets such that $ X \approx  N $ (as shown in the realistic examples for performance benchmarks). This lower ratio reduces the occ-RI-K speedup to $\sim \frac{2 N}{3 O}$, which is a bit smaller compared to standard energy calculations but still significant. For more detailed recommendations on the choice of basis set for core-focused property calculation, we refer the readers to an extensive discussion by Ireland and McKemmish \cite{ireland2023specialization}.

\section{Results}
\subsection{General computational settings}
We implemented the aforementioned algorithm in a development version of Q-Chem \cite{epifanovsky2021software}, which is now available in the public release version of Q-Chem 7.0. To examine the performance of our algorithm, we performed numerical accuracy and timing benchmarks using the state-of-the-art fully analytical implementation in ORCA 6.1.1 \cite{neese2025software} as a comparison. Production calculations were performed on the NERSC Perlmutter supercomputing system's CPU nodes (2 AMD EPYC Milan 7763 CPUs per node, with 64 physical cores per CPU, and 512 GB DDR4 memory on a node), with all jobs occupying the entire physical node to prevent interference on memory bandwidth from other jobs. 

For the analytical ORCA calculations, we follow the general suggestions from the original RIJCOSX-NMR literature \cite{stoychev2018self} to reduce numerical noise: the GTO-SCF convergence is controlled by the keyword ``VeryTightSCF'', and GIAO-CPSCF convergence is set to $10^{-7}$ ppm. If the RI approximation is applied, we apply the same treatment (i.e. RI-JK or RIJCOSX) for both zero-field GTO-SCF and GIAO-CPSCF. For the FD1 Q-Chem calculations, the SCF convergence for both zero-field GTO-SCF and field-on GIAO-SCF was set to $10^{-10}$ Hartree to achieve similar overall numerical accuracy. All calculations used the DIIS algorithm \cite{pulay1980convergence, pulay1982improved} to accelerate SCF convergence. All NMR shielding values are rounded to 3 decimal places to match ORCA's default printing style. More specific computational details for benchmarking numerical accuracy and computational performance will be explained in the corresponding sections. 

\subsection{Numerical accuracy of finite first derivative-based shieldings}
We first test the numerical accuracy of the finite first derivative-based (FD1) shieldings against the analytical implementation. Lutnaes, Teale, and coworkers performed systematic benchmark studies on magnetic molecular properties including the rotational g-tensor, magnetizability, NMR shielding constants, and nuclear spin-rotation constants \cite{lutnaes2009benchmarking, teale2013benchmarking}. Their test set contains 27 small molecules (we excluded O$_3$ as it is strongly correlated and thus unsuitable for DFT calculations) that span a wide range of different nuclei (unique shielding value count: 18 $^{1}$H, 2 $^{7}$Li, 17 $^{13}$C, 7 $^{15}$N, 11 $^{17}$O, 9 $^{19}$F, 2 $^{31}$P, 3 $^{33}$S), with a range of chemical environments. The geometries were optimized at the CCSD(T)/cc-pVTZ level and the reference shielding constants were at the CCSD(T)/CBS level  (extrapolated from aug-cc-pCVTZ and aug-cc-pCVQZ \cite{kendall1992electron, woon1995gaussian}). 
We use this dataset to perform FD1 vs analytical comparisons of numerical accuracy with Hartree-Fock theory and a rung-4 hybrid functional, TPSSh \cite{tao2003climbing, staroverov2003comparative}.

The following abbreviations are used to report error statistics: MSE for mean signed error, MAE for mean absolute error, RMSE for root mean square error, and MaxAE for max absolute error. Due to the greater experimental significance and natural abundance, results for $^{1}$H, $^{13}$C, and $^{19}$F are presented in the main text, while results for other elements are presented in the SI.

\subsubsection{Errors from finite magnetic field}
Here we compare the errors from the present FD1 finite magnetic field shieldings with the corresponding fully analytical result with the same treatment (i.e. exact JK build or RI approximated JK build with def2-universal-JK auxiliary basis \cite{weigend2008hartree}, for both the GTO-SCF and the GIAO response equations in ORCA). Q-Chem's RI-J + occ-RI-K results are exactly the same with RI-JK results (as analyzed in Section \ref{sec:giao-ri}), so we co-label them as ``RI''. Q-Chem's RI calculations and ORCA's RI-JK calculations both used the def2-universal-JKFit auxiliary basis set.

\begingroup
\scriptsize
\setlength{\tabcolsep}{6pt}
\renewcommand{\arraystretch}{1.0}

\setlength{\LTleft}{\fill}
\setlength{\LTright}{\fill}

\begin{longtable}{
  l l l
  S[table-format=-1.4]
  S[table-format=1.3]
  S[table-format=1.3]
  S[table-format=1.3]
}

\caption{Errors in the Hartree-Fock shielding constant (unit: ppm) calculated from the finite difference first derivative (FD1) scheme versus the corresponding fully analytical results, for $^{1}$H, $^{13}$C, and $^{19}$F}
\label{tab:fd1-errors-hcf} \\

\toprule
nuclear type & primary basis & computational settings
& {MSE} & {MAE} & {RMSE} & {MaxAE} \\
\midrule
\endfirsthead

\caption[]{Errors in the Hartree-Fock shielding constant (unit: ppm) calculated from the finite difference first derivative (FD1) scheme versus the corresponding fully analytical results, for $^{1}$H, $^{13}$C, and $^{19}$F, continued} \\

\toprule
nuclear type & primary basis & computational settings
& {MSE} & {MAE} & {RMSE} & {MaxAE} \\
\midrule
\endhead

\endfoot

\multirow{10}{*}{$^{1}$H (18 data points)}
& \multirow{7}{*}{pcSseg-2}
& exact, $1\times10^{-3}$ a.u. & 0.000 & 0.000 & 0.000 & 0.000 \\*
& & exact, $5\times10^{-4}$ a.u. & 0.000 & 0.000 & 0.000 & 0.000 \\*
& & exact, $1\times10^{-4}$ a.u. & 0.000 & 0.000 & 0.000 & 0.000 \\*
& & RI, $1\times10^{-3}$ a.u. & 0.000 & 0.000 & 0.000 & 0.000 \\*
& & RI, $5\times10^{-4}$ a.u. & 0.000 & 0.000 & 0.000 & 0.000 \\*
& & RI, $1\times10^{-4}$ a.u. & 0.000 & 0.000 & 0.000 & 0.000 \\*
\cline{2-7}

& \multirow{3}{*}{pcSseg-3}
& RI, $1\times10^{-3}$ a.u. & 0.000 & 0.000 & 0.000 & 0.000 \\*
& & RI, $5\times10^{-4}$ a.u. & 0.000 & 0.000 & 0.000 & 0.000 \\*
& & RI, $1\times10^{-4}$ a.u. & 0.000 & 0.000 & 0.000 & 0.000 \\*

\midrule

\multirow{10}{*}{$^{13}$C (17 data points)}
& \multirow{7}{*}{pcSseg-2}
& exact, $1\times10^{-3}$ a.u. & 0.001 & 0.001 & 0.001 & 0.002 \\*
& & exact, $5\times10^{-4}$ a.u. & 0.000 & 0.000 & 0.000 & 0.001 \\*
& & exact, $1\times10^{-4}$ a.u. & 0.000 & 0.000 & 0.000 & 0.001 \\*
& & RI, $1\times10^{-3}$ a.u. & 0.000 & 0.001 & 0.001 & 0.002 \\*
& & RI, $5\times10^{-4}$ a.u. & 0.000 & 0.000 & 0.000 & 0.001 \\*
& & RI, $1\times10^{-4}$ a.u. & 0.000 & 0.000 & 0.001 & 0.003 \\*
\cline{2-7}

& \multirow{3}{*}{pcSseg-3}
& RI, $1\times10^{-3}$ a.u. & 0.000 & 0.001 & 0.001 & 0.003 \\*
& & RI, $5\times10^{-4}$ a.u. & 0.000 & 0.000 & 0.000 & 0.001 \\*
& & RI, $1\times10^{-4}$ a.u. & 0.000 & 0.001 & 0.001 & 0.004 \\*

\midrule

\multirow{10}{*}{$^{19}$F (9 data points)}
& \multirow{7}{*}{pcSseg-2}
& exact, $1\times10^{-3}$ a.u. & 0.000 & 0.001 & 0.002 & 0.004 \\*
& & exact, $5\times10^{-4}$ a.u. & 0.000 & 0.001 & 0.001 & 0.002 \\*
& & exact, $1\times10^{-4}$ a.u. & 0.001 & 0.003 & 0.004 & 0.009 \\*
& & RI, $1\times10^{-3}$ a.u. & 0.000 & 0.001 & 0.004 & 0.009 \\*
& & RI, $5\times10^{-4}$ a.u. & 0.000 & 0.001 & 0.001 & 0.002 \\*
& & RI, $1\times10^{-4}$ a.u. & 0.003 & 0.006 & 0.009 & 0.022 \\*
\cline{2-7}

& \multirow{3}{*}{pcSseg-3}
& RI, $1\times10^{-3}$ a.u. & -0.001 & 0.002 & 0.002 & 0.004 \\*
& & RI, $5\times10^{-4}$ a.u. & 0.000 & 0.001 & 0.001 & 0.003 \\*
& & RI, $1\times10^{-4}$ a.u. & -0.006 & 0.013 & 0.020 & 0.049 \\*

\end{longtable}
\endgroup

In our finite field calculations, we choose 3 different magnetic field strengths, $1\times10^{-3}$ a.u., $5\times10^{-4}$ a.u., and $1\times10^{-4}$ a.u. The results show that the absolute finite field error is relatively small for all nuclear types at the tested levels, with typical RMSE values smaller than 0.010 ppm. The field strength of $5\times10^{-4}$ a.u. is optimal for minimizing the error with respect to the fully analytical reference for most of the nuclear types. At the optimal field strength of $5\times10^{-4}$ a.u., all RMSE values are $\le 0.003$ ppm, and no individual nuclear shielding differs from its corresponding analytical reference by more than 0.005 ppm. It is interesting to see that the optimal field strength is close to that being applied in practical experiments. A 100MHz benchtop NMR spectrometer corresponds to $10^{-5}$ a.u. field strength, while a 1GHz instrument corresponds to $10^{-4}$ a.u.
At similar field strengths, the $\hat{H}(\mathbf{B}^{(1)})$ we use is capable of capturing the correct physics without introducing significant high order magnetic response (due to the applied field being too large) or numerical differentiation noise (due to the applied field being too small for the finite numerical precision of the calculations).

We do observe a small element dependence in the numerical errors. For lighter nuclei such as $^{1}$H, $^{7}$Li, and $^{13}$C, most of the finite field shieldings agree with the reference to the third decimal place or 0.001 ppm at all 3 field strengths. The errors become a bit larger for heavier nuclei at the non-optimal field strengths $1\times10^{-3}$ a.u. or $1\times10^{-4}$ a.u., but remain about one order of magnitude smaller than experimental uncertainty ($\sim$ 0.1 ppm). The basis set dependence of the numerical errors is insignificant. 

\subsubsection{Errors from FD1 and the RI approximation versus RIJCOSX errors}
We next elect to test the combined overall errors from our use of FD1 scheme together with the RI approximation, with the reference defined as the fully analytical results with the exact (2e4c) JK build. For additional perspective, the errors from using the RIJCOSX or RI-JK approximations in the fully analytical calculations using ORCA are also presented. Again, the FD1 RI calculations (using our new implementation in Q-Chem) and the fully analytical RI-JK calculations (using ORCA) both use the def2-universal-JKFit auxiliary basis set (denoted as ``def2-JK''). For RIJCOSX calculations using ORCA, the default setting uses the smaller def2-universal-JFit (denoted as ``def2-J'') \cite{weigend2006accurate}. The errors from both methods are tested. The COSX grid is set to ``DefGrid3'' as originally recommended\cite{stoychev2018self}.

In general, the FD1 RI/def2-JK results reproduce analytical RI-JK/def2-JK error statistics very closely. This observation demonstrates that the finite field errors are usually much smaller than the errors associated with the RI approximation. Interestingly, we observe some small error cancellation between the finite field error and the RI error when using a slightly larger finite magnetic field strength, $\Delta B=1\times10^{-3}$ a.u., compared to the $5\times10^{-4}$ a.u. field strength that minimizes the finite field error against the fully analytical reference. This error cancellation is systematic across almost all nuclear types tested here, making the FD1 RI/def2-JK overall errors with $1\times10^{-3}$ a.u. applied field slightly smaller than those of analytical RI-JK/def2-JK relative to the fully analytical reference values with exact 2e4c integrals.

Numerical errors in the analytical RIJCOSX algorithm \cite{stoychev2018self} are associated with approximating J and/or K contributions to obtain the unperturbed density in GTO-SCF and the perturbed density in GIAO-CPSCF. When RIJCOSX was applied for both the GTO-SCF and GIAO-CPSCF, the largest error was caused by using the smaller def2-J auxiliary basis to calculate the Coulomb contribution to the unperturbed Fock matrix \cite{stoychev2018self}. We also observe similar numerical errors (Table \ref{tab:overall-errors-hcf}) with def2-J, and we find the errors are element-dependent. For lighter nuclei, the RMSE values of RIJCOSX/def2-J are small (e.g. $\sim$ 0.005 ppm for $^1$H, and $\sim$ 0.029 ppm for $^{13}$C). However, RIJCOSX/def2-J's RMSE values become much larger for heavier elements (e.g. $\sim$ 0.2-0.3 ppm for $^19$F, $>$ 0.4 ppm for $^{17}$O)
with some MaxAE values exceeding 1 ppm. Such errors are reduced by about one order of magnitude when the larger def2-JK auxiliary basis is used in RIJCOSX, reaching smaller errors than RI-JK/def2-JK. The remaining errors in RI-JK/def2-JK are partly due to the def2-JK auxiliary basis set not being designed to fit integrals from the pcS or pcSseg AO basis sets. Nonetheless, they are typically acceptable with $<$ 0.1 ppm RMSE values for all nuclei, except for $^{33}$S (RMSE $\sim$ 0.2 ppm with pcSseg-2 and $\sim$ 0.25 ppm with pcSseg-3).

\begingroup
\scriptsize
\setlength{\tabcolsep}{6pt}
\renewcommand{\arraystretch}{1.0}

\setlength{\LTleft}{\fill}
\setlength{\LTright}{\fill}

\begin{longtable}{
  l l l
  S[table-format=-1.4]
  S[table-format=1.3]
  S[table-format=1.3]
  S[table-format=1.3]
}

\caption{Overall combined errors in the Hartree-Fock shielding constant (unit: ppm) relative to the fully analytical results with exact JK build, for $^{1}$H, $^{13}$C, and $^{19}$F}
\label{tab:overall-errors-hcf} \\

\toprule
nuclear type & primary basis & computational settings
& {MSE} & {MAE} & {RMSE} & {MaxAE} \\
\midrule
\endfirsthead

\caption[]{Overall combined errors in the Hartree-Fock shielding constant (unit: ppm) relative to the fully analytical results with exact JK build, for $^{1}$H, $^{13}$C, and $^{19}$F, continued} \\

\toprule
nuclear type & primary basis & computational settings
& {MSE} & {MAE} & {RMSE} & {MaxAE} \\
\midrule
\endhead

\endfoot

\multirow{12}{*}{$^{1}$H (18 data points)}
& \multirow{6}{*}{pcSseg-2}
& FD1, RI/def2-JK, $1\times10^{-3}$ a.u. & 0.000 & 0.001 & 0.001 & 0.001 \\*
& & FD1, RI/def2-JK, $5\times10^{-4}$ a.u. & 0.000 & 0.001 & 0.001 & 0.001 \\*
& & FD1, RI/def2-JK, $1\times10^{-4}$ a.u. & 0.000 & 0.001 & 0.001 & 0.001 \\*
& & analytical, RIJCOSX/def2-J & -0.002 & 0.003 & 0.005 & 0.014 \\*
& & analytical, RIJCOSX/def2-JK & 0.001 & 0.001 & 0.001 & 0.003 \\*
& & analytical, RI-JK/def2-JK & 0.000 & 0.001 & 0.001 & 0.002 \\*
\cline{2-7}

& \multirow{6}{*}{pcSseg-3}
& FD1, RI/def2-JK, $1\times10^{-3}$ a.u. & 0.000 & 0.001 & 0.001 & 0.001 \\*
& & FD1, RI/def2-JK, $5\times10^{-4}$ a.u. & 0.000 & 0.001 & 0.001 & 0.001 \\*
& & FD1, RI/def2-JK, $1\times10^{-4}$ a.u. & 0.000 & 0.001 & 0.001 & 0.001 \\*
& & analytical, RIJCOSX/def2-J & -0.002 & 0.003 & 0.005 & 0.012 \\*
& & analytical, RIJCOSX/def2-JK & 0.001 & 0.001 & 0.001 & 0.002 \\*
& & analytical, RI-JK/def2-JK & 0.000 & 0.001 & 0.001 & 0.001 \\*

\midrule

\multirow{12}{*}{$^{13}$C (17 data points)}
& \multirow{6}{*}{pcSseg-2}
& FD1, RI/def2-JK, $1\times10^{-3}$ a.u. & -0.007 & 0.010 & 0.013 & 0.031 \\*
& & FD1, RI/def2-JK, $5\times10^{-4}$ a.u. & -0.007 & 0.010 & 0.013 & 0.032 \\*
& & FD1, RI/def2-JK, $1\times10^{-4}$ a.u. & -0.007 & 0.010 & 0.013 & 0.032 \\*
& & analytical, RIJCOSX/def2-J & 0.017 & 0.017 & 0.029 & 0.095 \\*
& & analytical, RIJCOSX/def2-JK & 0.007 & 0.007 & 0.012 & 0.030 \\*
& & analytical, RI-JK/def2-JK & -0.007 & 0.009 & 0.013 & 0.032 \\*
\cline{2-7}

& \multirow{6}{*}{pcSseg-3}
& FD1, RI/def2-JK, $1\times10^{-3}$ a.u. & -0.003 & 0.009 & 0.012 & 0.031 \\*
& & FD1, RI/def2-JK, $5\times10^{-4}$ a.u. & -0.003 & 0.010 & 0.012 & 0.031 \\*
& & FD1, RI/def2-JK, $1\times10^{-4}$ a.u. & -0.003 & 0.010 & 0.013 & 0.032  \\*
& & analytical, RIJCOSX/def2-J & 0.015 & 0.018 & 0.028 & 0.091 \\*
& & analytical, RIJCOSX/def2-JK & 0.006 & 0.006 & 0.009 & 0.021 \\*
& & analytical, RI-JK/def2-JK & -0.004 & 0.010 & 0.012 & 0.032 \\*

\midrule

\multirow{12}{*}{$^{19}$F (9 data points)}
& \multirow{6}{*}{pcSseg-2}
& FD1, RI/def2-JK, $1\times10^{-3}$ a.u. & 0.003 & 0.017 & 0.019 & 0.031 \\*
& & FD1, RI/def2-JK, $5\times10^{-4}$ a.u. & 0.003 & 0.016 & 0.018 & 0.031 \\*
& & FD1, RI/def2-JK, $1\times10^{-4}$ a.u. & 0.007 & 0.016 & 0.018 & 0.034 \\*
& & analytical, RIJCOSX/def2-J & 0.011 & 0.092 & 0.126 & 0.261 \\*
& & analytical, RIJCOSX/def2-JK & -0.011 & 0.019 & 0.031 & 0.078 \\*
& & analytical, RI-JK/def2-JK & 0.004 & 0.016 & 0.018 & 0.031 \\*
\cline{2-7}

& \multirow{6}{*}{pcSseg-3}
& FD1, RI/def2-JK, $1\times10^{-3}$ a.u. & 0.010 & 0.025 & 0.032 & 0.062 \\*
& & FD1, RI/def2-JK, $5\times10^{-4}$ a.u. & 0.010 & 0.025 & 0.031 & 0.058 \\*
& & FD1, RI/def2-JK, $1\times10^{-4}$ a.u. & 0.005 & 0.032 & 0.041 & 0.068 \\*
& & analytical, RIJCOSX/def2-J & 0.008 & 0.091 & 0.126 & 0.235 \\*
& & analytical, RIJCOSX/def2-JK & -0.003 & 0.017 & 0.023 & 0.058 \\*
& & analytical, RI-JK/def2-JK & 0.010 & 0.025 & 0.031 & 0.058 \\*

\end{longtable}
\endgroup

For NMR shielding calculations with DFT, the use of finite numerical quadrature grids for semi-local exchange and correlation is another source of numerical error. We tested the combined errors of Q-Chem and ORCA calculations from the rung-4 TPSSh functional \cite{tao2003climbing, staroverov2003comparative}, which gave good results for $^{1}$H and $^{13}$C chemical shifts
in extensive benchmark tests\cite{schattenberg2021extended}. The field-dependent, gauge-invariant version $\tau$ treatment of Maximoff and Scuseria \cite{maximoff2004nuclear} is used for the meta GGA part of TPSSh for both Q-Chem and ORCA.

The statistics (presented in the SI) establish that errors from using the pruned standard grid SG-2 \cite{dasgupta2017standard} (constructed from a parent grid with 75 radial spheres and 302 angular Lebedev grid points) are relatively small for the lighter nuclei ($\sim$ 0.006 ppm RMSE for $^1$H and $\sim$ 0.03 ppm RMSE for $^{13}$C). However, for heavier nuclei, the errors from using the SG-2 grid are larger, with RMSE values $>$ 0.12 ppm for $^{15}$N, $>$ 0.08 ppm for $^{17}$O, $>$ 0.07 ppm for $^{19}$F, $>$ 0.35 ppm for $^{31}$P. Such errors nearly vanish when the unpruned (75,302) grid (parent grid of SG-2) is used for all nuclear types, bringing the numerical errors in Q-Chem's RI-based TPSSh results very close to ORCA's RI-JK/def2-JK. For normal total energy and force calculations, the pruning procedure deletes ``unnecessary'' grid points, with little or no pruning in the valence region. Nevertheless, for NMR shieldings (and other core properties), pruning angular grid points from the core region limits the ability of the grid to describe the density and its changes close to the nucleus, leading to a relatively large error in the final observable.

We also observe similar RIJCOSX/def2-J errors for HF and TPSSh, as both are dominated by the Coulomb contribution to the unperturbed Fock matrix. For TPSSh, ORCA's RIJCOSX/def2-JK is slightly less numerically accurate than RI-JK for most elements. The small fraction (10$\%$) of exact exchange in TPSSh dilutes the contribution of exchange to the overall errors. As supported by existing benchmarks\cite{stoychev2018self}, we expect analytical RI-JK/def2-JK and its FD1 alternative to be numerically more accurate than analytical RIJCOSX/def2-JK for other typical global hybrid functionals with higher fractions (e.g. $\sim 20 \% - 50 \%$) of exact exchange.

Overall, the errors due to finite difference and finite magnetic field are insignificant once the field strength is properly chosen, and is probably negligible for routine applications. While the numerical errors with RIJCOSX/def2-J in Hartree-Fock and hybrid DFT are still typically smaller compared to other sources of error (such as the error from the computational level of theory, basis set/grid incompleteness, and ignoring vibrational averaging), for heavier elements, the RMSEs start to approach or exceed experimental uncertainty. Thus we recommend caution when using RIJCOSX/def2-J for elements heavier than $^{13}$C. Using unpruned angular grids in DFT-based NMR shielding calculations is also recommended.

\subsection{Timing and parallel efficiency benchmarks on realistic systems}
We next examine timings for NMR shielding calculations of benzene, caffeine, aspartame (widely used artificial sweetener, also a dipeptide), ATP, cholesterol, and lipitor (brand name of atorvastatin, a statin medication used to treat high cholesterol and prevent heart attacks), in ascending order of system size. The Cartesian coordinates of benzene and caffeine are obtained from the pre-built structures in IQmol, and the rest are obtained from the first 3D conformation on the PubChem online database \cite{kim2025pubchem} (also attached in the SI). The computational level is chosen to be HF/pcSseg-3; very similar timings would be obtained with global hybrid functionals. The wall time is for the entire calculation to obtain the final shielding, and parallelization efficiency is defined as the ratio of the speedup relative to the serial wall time for the same method and the number of processors used in the parallel calculation.

For the Q-Chem calculations, we employed the FD1 algorithm to calculate the NMR shielding. Following the tests above, the finite magnetic field strength was set to $1\times10^{-3}$ a.u. to minimize the overall error relative to the shielding numbers from fully analytical second derivatives with exact JK. The RI-J + occ-RI-K algorithm were used for both the zero-field GTO-SCF and field-on GIAO-SCF, with intermediate 2e3c integrals computed once for each field strength and stored in memory.

For the ORCA calculations, the NMR shieldings were calculated analytically with the same memory as the corresponding Q-Chem calculation. RIJCOSX was used with the smaller def2-J auxiliary basis for the RI-J part and the ``DefGrid3'' for the COSX part. Although the numerical errors of RIJCOSX/def2-J on some heavier nuclei are relatively large (in contrast to that of RIJCOSX/def2-JK), the errors on the most chemically important nuclei, $^{1}$H and $^{13}$C, are still acceptable and much smaller than other error sources. We also selected RIJCOSX/def2-J because it is ORCA's default.
Changing to a smaller finite field strength $5\times10^{-4}$ for the FD1 calculations (field-on MO coefficients are closer to the zero-field solution, so the expensive GIAO-SCF takes fewer iterations to converge), or changing to the larger def2-JK auxiliary basis for RIJCOSX would push the performance comparison slightly in favor of the FD1 algorithm. 

\subsubsection{Single-node performance}
We first present single-node timing benchmarks with the 3 smaller molecules: benzene, caffeine, and aspartame. Since all computational resources are on the same node, the intermediate modules in FD1 shielding calculations are done in a purely sequential manner (with shared-memory OpenMP parallelization), i.e., zero-field GTO-SCF, then GIAO-SCF at $(+\Delta B, 0, 0)$, then GIAO-SCF at $(0, +\Delta B, 0)$, then GIAO-SCF at $(0,0,+\Delta B)$, then finally the finite difference step to get the full system shieldings. All 3 GIAO-SCF calculations still take the converged zero-field MO coefficients as the initial guess, and all GTO-SCF and GIAO-SCF calculations apply OpenMP parallelization. The fully analytical calculations apply MPI parallelization only, as this is ORCA's design.

\begin{figure}[H]
    \centering

    \begin{subfigure}{0.95\textwidth}
        \centering
        \includegraphics[width=\textwidth]{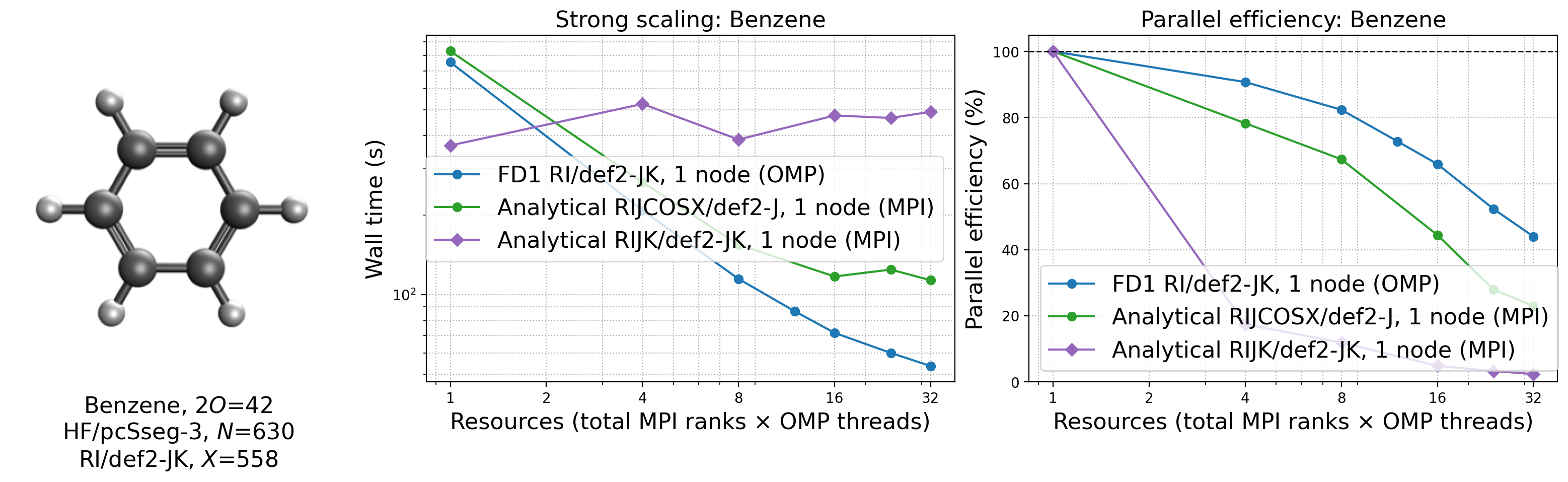}
        \label{fig:benzene}
    \end{subfigure}

    \begin{subfigure}{0.95\textwidth}
        \centering
        \includegraphics[width=\textwidth]{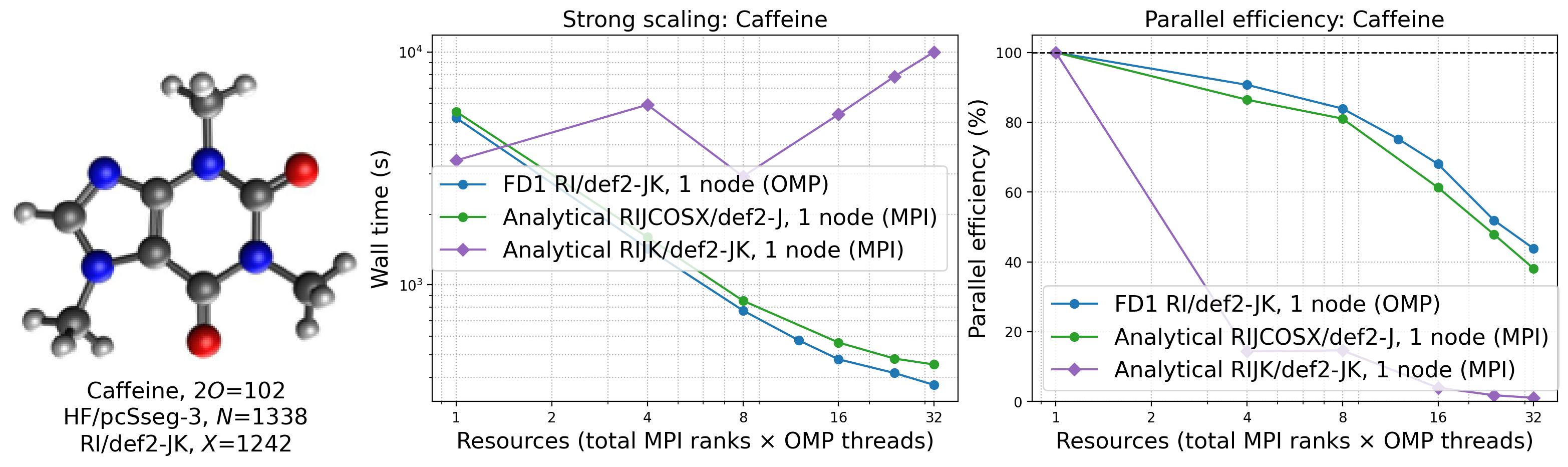}
        \label{fig:caffeine}
    \end{subfigure}

    \begin{subfigure}{0.95\textwidth}
        \centering
        \includegraphics[width=\textwidth]{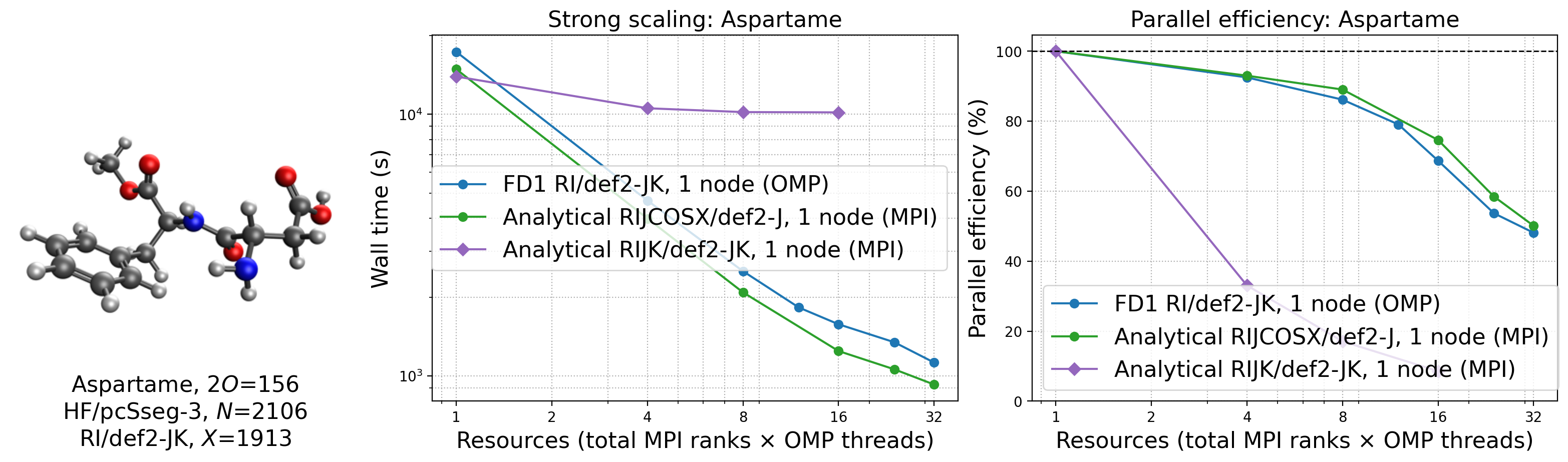}
        \label{fig:aspartame}
    \end{subfigure}

    \caption{Single-node performance for the 3 smaller systems: benzene, caffeine, and aspartame. The number of electrons ($2O$), AOs ($N$) (with the pcSseg-3 basis set), and auxiliary basis functions ($X$) (with the def2-JK auxiliary basis set used in the FD1 calculations) are listed beside the strong scaling plots.}
    \label{fig:single-node}
\end{figure}

For NMR shielding calculations on relatively small systems (for instance, less than 100 electrons and 1000 basis functions), RI-JK is reported to be faster than RIJCOSX \cite{stoychev2018self}; therefore we also perform RI-JK calculations (with def2-JK auxiliary basis set) on the 3 small systems. Here in the strong scaling performance plots of Figure \ref{fig:single-node}, we also observe faster serial RI-JK wall time for all 3 systems, indicating that the baseline computational effort needed for RI-JK based CPSCF is smaller than RIJCOSX. However, the parallel efficiency of RI-JK consistently decays very fast with increasing resources. Starting from 4 physical cores, RIJCOSX's walltime is significantly faster than that of RI-JK, and the performance gap between analytical RI-JK and RIJCOSX becomes much larger for larger systems due to the steeper scaling. Therefore, we focus on the comparison between FD1 with RI-J + occ-RI-K and analytical RIJCOSX results.

For benzene and caffeine, FD1 RI-J + occ-RI-K exhibits a smooth and monotonic reduction in the wall time, while analytical RIJCOSX shows non-monotonic wall clock behavior for benzene. In these 2 cases, FD1 consistently outperforms analytic on both wall time and parallel efficiency, with $\sim 110 \%$ performance advantage on benzene and $\sim 22 \%$ on caffeine with 32 physical cores. With aspartame, RIJCOSX becomes more favorable as a result of its lower scaling with system size (practically $\mathcal{O}(N^2)$ to $\mathcal{O}(N^3)$ for COSX vs $\mathcal{O}(N^4)$ for occ-RI-K). 
FD1 RI-J + occ-RI-K results still closely follow the ORCA results with almost the same parallel efficiency ($\sim 50 \%$) at 32 physical cores.

\subsubsection{Multi-node performance}
\begin{figure}[H]
    \centering

    \begin{subfigure}{0.95\textwidth}
        \centering
        \includegraphics[width=\textwidth]{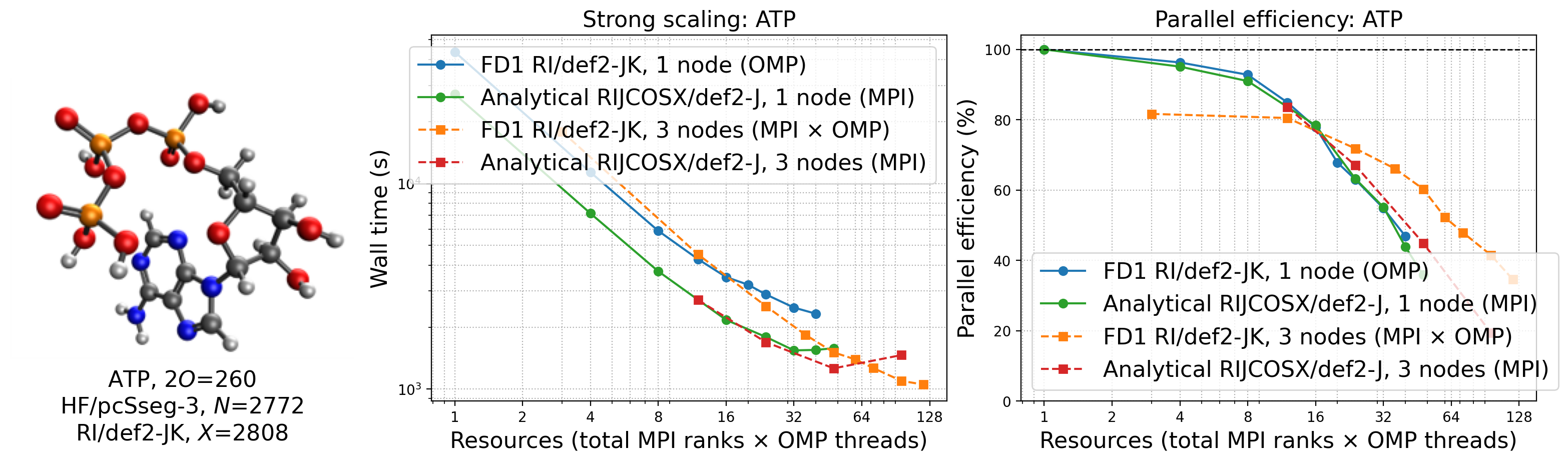}
        \label{fig:benzene}
    \end{subfigure}

    \begin{subfigure}{0.95\textwidth}
        \centering
        \includegraphics[width=\textwidth]{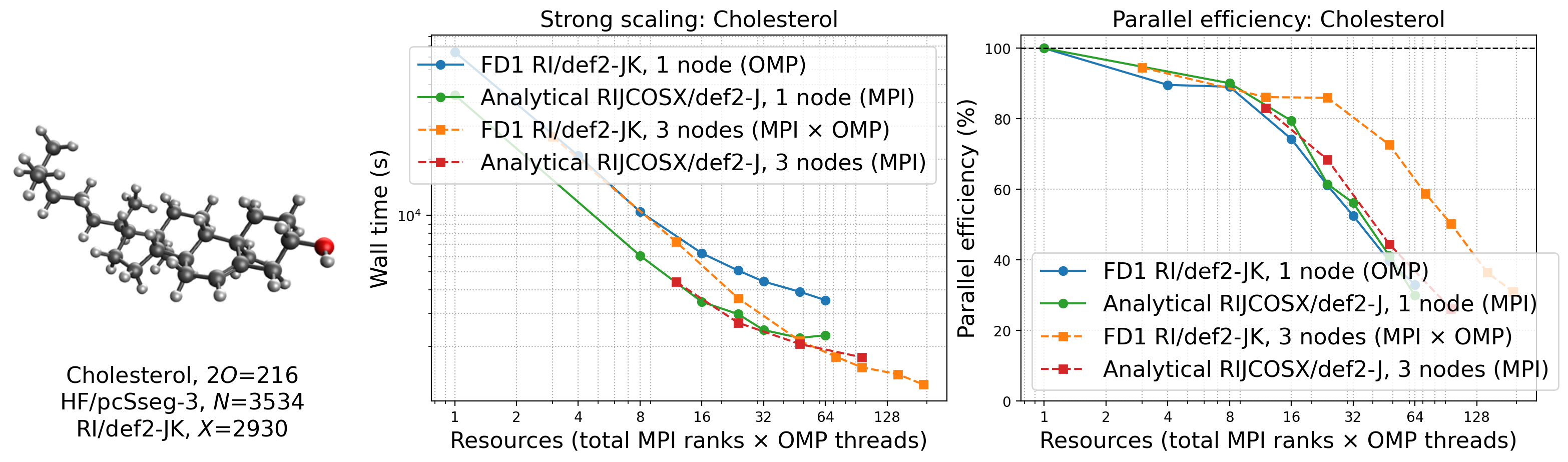}
        \label{fig:caffeine}
    \end{subfigure}

    \begin{subfigure}{0.95\textwidth}
        \centering
        \includegraphics[width=\textwidth]{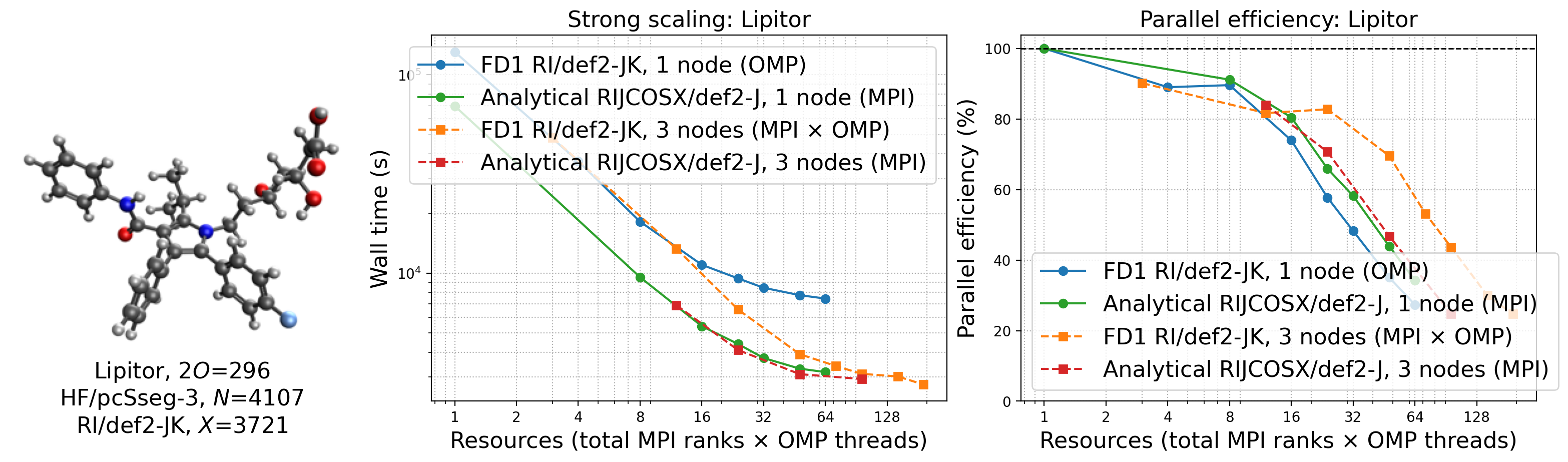}
        \label{fig:aspartame}
    \end{subfigure}

    \caption{Multi-node performance for the 3 larger systems: ATP, cholesterol, and lipitor. The number of electrons ($2O$), AOs ($N$) (with the pcSseg-3 basis set), and auxiliary basis functions ($X$) (with the def2-JK auxiliary basis set used in the FD1 calculations) are listed beside the strong scaling plots.}
    \label{fig:multi-node}
\end{figure}

In Figure \ref{fig:multi-node} we present the timing benchmarks with 3 larger molecules, ATP, cholesterol, and lipitor, including both the single-node runs (solid curves) and multi-node runs (dashed curves). Again, due to the more favorable scaling,  RIJCOSX shows consistently faster wall time starting from serial to multiple physical cores in single-node runs. Nevertheless, for multi-node runs, a new source of performance gain is introduced from more memory bandwidth and more cores on multiple physical nodes. With its natural dual-level parallelism, the multi-node performance of the FD1 algorithm significantly benefits by efficiently leveraging MPI and OpenMP hybrid parallelization. On the other hand, the RIJCOSX wall time gradually plateaus and reaches a minimum due to MPI communication overhead outweighing the gains from more computational resources. The FD1 wall time still shows a monotonic reduction with increasing resources, with better ($\sim 10-20 \%$ greater) parallel efficiency relative to the analytical run at 3$\times$8 resources (that is, 3 MPI ranks on 3 physical nodes with 8 OMP threads per physical node for FD1, or 3 physical nodes x 8 MPI ranks per physical node for analytical) and beyond. The much better parallelization efficiency brings the FD1 wall time down relative to the analytical, with a crossover in wall time at 3$\times$24 resources for ATP and cholesterol.

In Table \ref{tab:parallel}, the parallel efficiency advantage of the MPI/OMP strategy in our FD1 algorithm becomes more obvious. At the same total resource count of 24, the 3$\times$8 MPI/OMP configuration is $\sim 12 \%$, $\sim29 \%$, and $\sim30 \%$ faster than the 1$\times$24 configuration for ATP, cholesterol, and Lipitor, respectively. The corresponding improvements in analytical are only $\sim6 \%$, $\sim10 \%$, and $\sim7 \%$, indicating that our FD1 implementation benefits substantially more from distributing the workload across multiple MPI ranks. This difference is even clearer when the resources are increased from 1$\times$32 to 3$\times$32. Our implementation achieves speedups of 2.27, 2.87, and 2.71 for the three systems. In comparison, the analytical run achieves speedups of only 1.05, 1.40, and 1.28. Consequently, although the FD1 implementation is $\sim 57 - 91 \%$ slower than analytical RIJCOSX at 1$\times$8, it becomes $\sim 25 \%$ faster for ATP and $\sim 12 \%$ faster for cholesterol at 3$\times$32. For lipitor, the performance gap is reduced from $\sim 91 \%$ at 1$\times$8 to only $\sim 6 \%$ at 3$\times$32. These results demonstrate that the hybrid MPI/OpenMP implementation of FD1 converts additional distributed computing resources into wall-time reductions much more effectively than a conventional analytical implementation, allowing a performance advantage to emerge at larger resource counts.

\begin{table}[h!]
\begin{tabular}{@{}cccc@{}}
  & \shortstack{ATP \\ \textbf{$N=2772$}}   & \shortstack{Cholesterol \\ \textbf{$N=3534$}} & \shortstack{Lipitor \\ \textbf{$N=4107$}}  \\ 
\midrule
\shortstack{MPI x OMP \\ 1 x 8} & \shortstack{FD1: 5883 \\ Analytical: 3739} & \shortstack{FD1: 10471 \\ Analytical: 6098} & \shortstack{FD1: 18213 \\ Analytical: 9554} \\
\midrule
\shortstack{MPI x OMP \\ 3 x 8} & \shortstack{FD1: 2533 \\ Analytical: 1692} & \shortstack{FD1: 3617 \\ Analytical: 2677} & \shortstack{FD1: 6572 \\ Analytical: 4106} \\
\midrule
\shortstack{MPI x OMP \\ 1 x 24} & \shortstack{FD1: 2887 \\ Analytical: 1793} & \shortstack{FD1: 5087 \\ Analytical: 2976} & \shortstack{FD1: 9426 \\ Analytical: 4404} \\
\midrule
\shortstack{MPI x OMP \\ 1 x 32} & \shortstack{FD1: 2490 \\ Analytical: 1542} & \shortstack{FD1: 4437 \\ Analytical: 2445} & \shortstack{FD1: 8451 \\ Analytical: 3741} \\
\midrule
\shortstack{MPI x OMP \\ 3 x 32} & \shortstack{FD1: 1096 \\ Analytical: 1464} & \shortstack{FD1: 1548 \\ Analytical: 1751} & \shortstack{FD1: 3114 \\ Analytical: 2934} \\
\bottomrule
\caption {Wall time (in seconds) comparison between analytical (RIJCOSX/def2-J) and FD1 (RI-J + occ-RI-K with def2-JK) algorithms for chemical shielding using the specified computing resources (MPI ranks times OMP threads) for 3 larger systems, ATP, cholesterol, and lipitor.
\label{tab:parallel}} 
\end{tabular}
\end{table}

\subsection{Application: structure determination of a natural product}

Finally, we employ our efficient FD1 algorithm for NMR shielding on a chemical application. Hexacylinol is a complex bioactive natural product originally isolated as a fungus metabolite from dead betula woods in Siberia \cite{schlegel2002hexacyclinol}. While its anti-cancer properties still attract scientific interest, its structure determination has been a historical controversy until its total synthesis by John Porco, Jr. in 2006 \cite{porco2006total}. DFT-based NMR shielding calculations also played a crucial role in confirming the correctness of the revised structure \cite{rychnovsky2006predicting, saielli2009can}.

\begin{figure}[H]
    \centering

    \begin{subfigure}{0.95\textwidth}
        \centering
        \includegraphics[width=\textwidth]{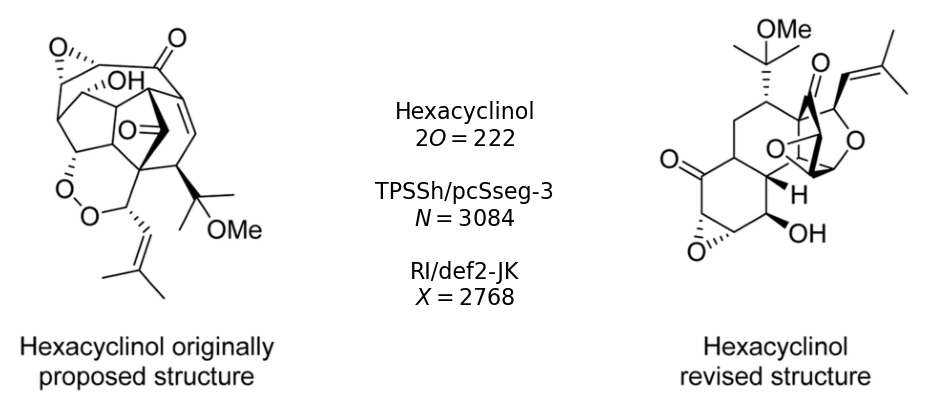}
        \label{fig:hexacyclinol_structure}
    \end{subfigure}

    \begin{subfigure}{0.95\textwidth}
        \centering
        \includegraphics[width=\textwidth]{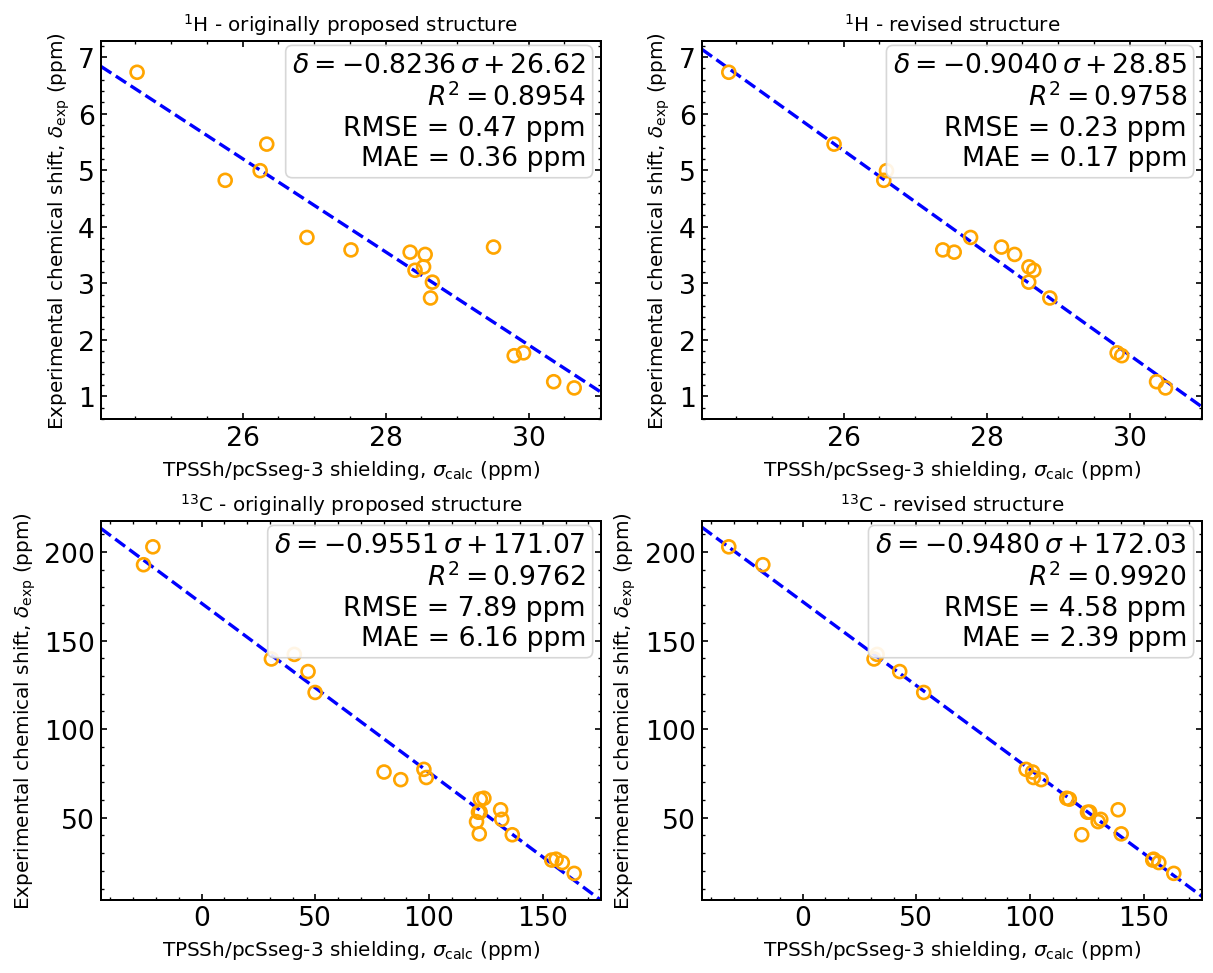}
        \label{fig:hexaxyclionl_fit}
    \end{subfigure}

    \caption{Linear fit based on TPSSh/pcSseg-3 calculations against hexacylinol experimental chemical shifts for $^1$H and $^{13}$C, with the originally proposed structure (top left) and the revised structure (top right) adapted from a recent review by de Albuquerque et al. \cite{de2024we} under CC-BY 4.0 license}

    \label{fig:hexacyclinol}
\end{figure}

Here we calculate the hexacylinol NMR shieldings for the originally proposed and the revised structures using our new FD1 algorithm in Q-Chem. We started from the structures reported previously\cite{saielli2009can} (optimized at B3LYP/6-31G(d,p) \cite{hehre1972self, hariharan1973influence} level including two conformers of the revised structure), and then performed further geometry optimization with TPSSh-D3(BJ) \cite{grimme2010consistent, grimme2011effect} /def2-TZVPP, as recommended in an extensive benchmark of DFT optimized equilibrium structures \cite{karton2021evaluation}. Frequency calculations at the same level were performed to ensure that both structures are true minima (i.e. without imaginary frequencies). Single point electronic energy calculations were performed at the  $\omega$B97M-V \cite{mardirossian2016omegab97m} /def2-QZVPPD level, and the free energy as well as the corresponding Boltzmann factor of the two conformers of the revised structure were calculated from the $\omega$B97M-V/def2-QZVPPD electronic energy plus the zero-point vibrational energy and non-thermo energy corrections at the TPSSh-D3(BJ)/def2-TZVPP level. Finally, the NMR shielding calculations are done at TPSSh/pcSseg-3 level, using the RI-J + occ-RI-K approximated Fock build with def2-JK auxiliary basis. The $^1$H and $^{13}$C chemical shifts are obtained from a linear fit against experimental values shown in Figure \ref{fig:hexacyclinol}. As expected, the revised structure shows significantly smaller RMSE and MAE for both $^1$H and $^{13}$C with higher $R^{2}$, demonstrating the utility of the FD1 algorithm for natural product structure determination.

\section{Conclusions}
In this work, we have designed and implemented a very efficient algorithm that we called the FD1 approach for density functional theory calculations of NMR shieldings. FD1 performs finite first differences on analytic first derivatives of DFT energies with respect to nuclear magnetic moments, in the existence of finite magnetic fields. We prove that only 3 finite field calculations using forward differences are necessary to obtain all closed shell magnetic shieldings with the same second order accuracy as central differences. The finite field shielding calculations are performed in complex arithmetic using GIAOs, and are accelerated by complex extensions of resolution of the identity (or density fitting) methods and dual-level MPI/OMP parallelization.  

Thorough numerical accuracy benchmarks establish that errors from finite field differentiation are much smaller than other error sources and can be neglected for routine applications. To minimize numerical errors, we recommend that the def2-JK auxiliary basis be applied with RIJCOSX, and that unpruned grids are applied with DFT. In performance benchmarks on realistic chemical systems of different sizes, our FD1 algorithm implemented in Q-Chem achieves comparable or better performance than the state-of-the-art analytic implementation in ORCA 6.1.1 on small to intermediate systems, while for the largest system with more than 4000 basis functions, our parallelization efficiency remains very strong.

In summary, our work reintroduces a simple yet powerful scheme for molecular magnetic properties calculations, which is extended with error analysis, and adapted to naturally exploit dual-level MPI/OpenMP parallelization. That approach is transferable for other molecular properties calculated in finite first differences, or by analogy, distributing the individual response equations on different MPI ranks for the full analytic response formalism. 

The specifics of our FD1 implementation can potentially be further enhanced in future work. For example, RI-SCF methods can benefit from density compression \cite{lara2026algorithm} as well as the semi-numerical K build \cite{laqua2020highly} that originates from chain-of-spheres exchange (COSX) \cite{neese2009efficient} and pseudospectral predecessors \cite{friesner1985solution}. Also, our algorithm was implemented for DFT calculations here, but there is considerable potential to extend the FD1 to correlated (non-variational) wavefunction theory. Other enhancements to efficiency and/or accuracy are also possible, such as using locally dense basis sets \cite{liang2023efficient}, local correlation \cite{wang2025more} or embedding \cite{shee2024static}, and inclusion of relativistic effects at different levels \cite{vicha2020relativistic}. We hope to report on some of these possibilities in future work. 

\section*{Conflicts of Interest}
MH-G is a part-owner of Q-Chem Inc, whose software was used as the implementation platform for the algorithm reported in this work.

\begin{acknowledgement}
This work was supported by the Director, Office of Basic Energy Sciences, Chemical Sciences, Geosciences, and Biosciences Division of the U.S. Department of Energy, under Contract No. DE-AC02-05CH11231. This research used resources of the National Energy Research Scientific Computing Center (NERSC), a U.S. Department of Energy Office of Science User Facility located at Lawrence Berkeley National Laboratory, operated under Contract No. DE-AC02-05CH11231. We would like to thank Dr. Henryk Laqua and Dr. Xintian Feng for helpful discussions.
\end{acknowledgement}







\section*{Appendix: GIAO partial derivative integrals}
With the $\hat{H}( \boldsymbol{m_{K}}^{(1)}, \boldsymbol{B}^{(1)})$ Hamiltonian in Section \ref{sec:hamiltonian}, partial derivative integral with respect to $\boldsymbol{m_{K}}$ in GIAO basis only contains contributions from the direct derivatives of the $\boldsymbol{m_{K}}$ dependent operators, namely paramagnetic spin-orbit (PSO) and diamagnetic shielding (DS), 

\begin{equation}
\begin{aligned}
    \frac{\partial h_{\mu \nu} \left( \mathbf{B} \right)}{\partial m_{K, \alpha}} & = \bra{ \omega_{\mu} \left( \mathbf{B} \right) } \frac{\partial \hat{H}( \boldsymbol{m_{K}}^{(1)}, \boldsymbol{B}^{(1)}) }{ \partial m_{K, \alpha}} \ket{ \omega_{\nu} \left( \mathbf{B} \right) } \\ 
    & = \bra{ \omega_{\mu} \left( \mathbf{B} \right) } \frac{\partial \hat{h}^{PSO} }{ \partial m_{K, \alpha}} \ket{ \omega_{\nu} \left( \mathbf{B} \right) } + \bra{ \omega_{\mu} \left( \mathbf{B} \right) } \frac{\partial \hat{h}^{DS} }{ \partial m_{K, \alpha}} \ket{ \omega_{\nu} \left( \mathbf{B} \right) }
\end{aligned}
\end{equation}

\begin{equation}
    \hat{h}^{PSO} = \alpha^2 \sum_{i} \sum_{K} \frac{\boldsymbol{m_{K}} \cdot \boldsymbol{L_{iK}}}{r_{iK}^3}
\end{equation}

\begin{equation}
    \hat{h}^{DS} = \frac{\alpha^2}{2} \sum_{i} \sum_{K} \frac{ (\boldsymbol{B} \cdot \boldsymbol{m_{K}}) (\boldsymbol{r_{i} \cdot \boldsymbol{r_{iK}}}) - (\boldsymbol{B} \cdot \boldsymbol{r_{iK})(\boldsymbol{r_{i} \cdot \boldsymbol{m_{K}}}}) }{r_{iK}^3}
\end{equation}

Here $\alpha$ denotes the fine structure constant. We evaluate the partial derivative integrals based on field integral (FI) building blocks with shifts on angular momenta. The fundamental integral kernels with GIAO basis are evaluated using the Taketa-Huzinaga-O-ohata (THO66) algorithm \cite{taketa1966gaussian} and its extension to molecular property integrals by Matsuoka \cite{matsuoka1971gaussian},
\begin{equation}
    FI_{\mu \nu}^{x} = \bra{\omega_{\mu}} \frac{x_K}{r^3_{iK}} \ket{\omega_{\nu}}
    \qquad
    FI_{\mu \nu}^{y} = \bra{\omega_{\mu}} \frac{y_K}{r^3_{iK}} \ket{\omega_{\nu}}
    \qquad
    FI_{\mu \nu}^{z} = \bra{\omega_{\mu}} \frac{z_K}{r^3_{iK}} \ket{\omega_{\nu}}
\end{equation}

Field integrals with angular momenta shifts are expressed as follows,
\begin{equation}
    FI_{\mu \nu}^{x}(l_{\mu} +1) = \bra{\omega_{\mu} (l_{\mu} +1)} \frac{x_K}{r^3_{iK}} \ket{\omega_{\nu}}
    \qquad
    FI_{\mu \nu}^{y}(m_{\nu} +1) = \bra{\omega_{\mu}} \frac{y_K}{r^3_{iK}} \ket{\omega_{\nu} (m_{\nu} +1)}
    \qquad
\end{equation}

where $l, m, n$ are the polynomial powers correspond to the Cartesian $x, y, z$ components of the GIAO basis function respectively.  

After some straightforward yet cumbersome algebra, we arrive at the following expression to evaluate the desired GIAO partial derivative integrals, where $\alpha_{\mu}$ and $\alpha_{\nu}$ are used to denote the Gaussian exponent for orbitals $\mu$ and $\nu$ respectively.

For partial derivatives of PSO integral,
\begin{equation}
\begin{aligned}
    \bra{\omega_{\mu}} \frac{\partial \hat{h}^{PSO}} {\partial m_{K, x} } \ket{\omega_{\nu}}
    = & -i \cdot \alpha^2
    \left[
    n_{\nu} FI_{\mu \nu}^{y} (n_{\nu}-1)
    - i \cdot \frac{ \chi_{N, z} }{2} FI_{\mu \nu}^{y}
    - 2 \alpha_{\nu} FI_{\mu \nu}^{ y} (n_{\nu} + 1) \right. \\
    & \left. \qquad \qquad
    - m_{\nu} FI_{\mu \nu}^{z} (m_{\nu} - 1)
    + i \cdot \frac{ \chi_{N, y} }{2} FI_{\mu \nu}^{z}
    + 2 \alpha_{\nu} FI_{\mu \nu}^{z} (m_{\nu} + 1)
    \right]
\end{aligned}
\end{equation}

\begin{equation}
\begin{aligned}
    \bra{\omega_{\mu}} \frac{\partial \hat{h}^{PSO}} {\partial m_{K, y} } \ket{\omega_{\nu}}
    = & -i \cdot \alpha^2
    \left[
    l_{\nu} FI_{\mu \nu}^{z} (l_{\nu}-1)
    - i \cdot \frac{ \chi_{N, x} }{2} FI_{\mu \nu}^{z}
    - 2 \alpha_{\nu} FI_{\mu \nu}^{z} (l_{\nu} + 1) \right. \\
    & \left. \qquad \qquad
    - n_{\nu} FI_{\mu \nu}^{x} (l_{\nu} - 1)
    + i \cdot \frac{ \chi_{N, x} }{2} FI_{\mu \nu}^{x}
    + 2 \alpha_{\nu} FI_{\mu \nu}^{x} (n_{\nu} + 1)
    \right]
\end{aligned}
\end{equation}

\begin{equation}
\begin{aligned}
    \bra{\omega_{\mu}} \frac{\partial \hat{h}^{PSO}} {\partial m_{K, z} } \ket{\omega_{\nu}}
    = & -i \cdot \alpha^2
    \left[
    m_{\nu} FI_{\mu \nu}^{x} (m_{\nu}-1)
    - i \cdot \frac{ \chi_{N, y} }{2} FI_{\mu \nu}^{x}
    - 2 \alpha_{\nu} FI_{\mu \nu}^{x} (m_{\nu} + 1) \right. \\
    & \left. \qquad \qquad
    - l_{\nu} FI_{\mu \nu}^{y} (l_{\nu} - 1)
    + i \cdot \frac{ \chi_{N, z} }{2} FI_{\mu \nu}^{y}
    + 2 \alpha_{\nu} FI_{\mu \nu}^{y} (l_{\nu} + 1)
    \right]
\end{aligned}
\end{equation}

where the intermediate vector quantity $\boldsymbol{ \chi_{N} }$ is defined as follows (subscripts $x, y, z$ indicate the corresponding component),
\begin{equation}
    \boldsymbol{ \chi_{N} } = \boldsymbol{B} \times \boldsymbol{R}_{N} = (\chi_{N,x} \ , \chi_{N,y} \ , \chi_{N,z})
\end{equation}

For partial derivatives of DS integral,
\begin{equation}
\begin{aligned}
    \bra{\omega_{\mu}} \frac{\partial \hat{h}^{DS}} {\partial m_{K, x} } \ket{\omega_{\nu}}
    = & \frac{ \alpha^2 }{2} \left[
    B_{x} FI_{\mu \nu}^{y} (m_{\nu} + 1) 
    - B_{y} FI_{\mu \nu}^{y} (l_{\nu} + 1)
    + (B_{x} N_{y} - B_{y} N_{x}) FI_{\mu \nu}^{y} \right. \\
    & \left. \qquad
    + B_{x} FI_{\mu \nu}^{z} (n_{\nu} + 1) 
    - B_{z} FI_{\mu \nu}^{z} (l_{\nu} + 1)
    + (B_{x} N_{z} - B_{z} N_{x}) FI_{\mu \nu}^{z}
    \right]
\end{aligned}
\end{equation}

\begin{equation}
\begin{aligned}
    \bra{\omega_{\mu}} \frac{\partial \hat{h}^{DS}} {\partial m_{K, y} } \ket{\omega_{\nu}}
    = & \frac{ \alpha^2 }{2} \left[
    B_{y} FI_{\mu \nu}^{x} (l_{\nu} + 1) 
    - B_{x} FI_{\mu \nu}^{x} (m_{\nu} + 1)
    + (B_{y} N_{x} - B_{x} N_{y}) FI_{\mu \nu}^{x} \right. \\
    & \left. \qquad
    + B_{y} FI_{\mu \nu}^{z} (n_{\nu} + 1) 
    - B_{z} FI_{\mu \nu}^{z} (m_{\nu} + 1)
    + (B_{y} N_{z} - B_{z} N_{y}) FI_{\mu \nu}^{z}
    \right]
\end{aligned}
\end{equation}

\begin{equation}
\begin{aligned}
    \bra{\omega_{\mu}} \frac{\partial \hat{h}^{DS}} {\partial m_{K, z} } \ket{\omega_{\nu}}
    = & \frac{ \alpha^2 }{2} \left[
    B_{z} FI_{\mu \nu}^{x} (l_{\nu} + 1) 
    - B_{x} FI_{\mu \nu}^{x} (n_{\nu} + 1)
    + (B_{z} N_{x} - B_{x} N_{z}) FI_{\mu \nu}^{x} \right. \\
    & \left. \qquad
    + B_{z} FI_{\mu \nu}^{y} (m_{\nu} + 1) 
    - B_{y} FI_{\mu \nu}^{y} (n_{\nu} + 1)
    + (B_{z} N_{y} - B_{y} N_{z}) FI_{\mu \nu}^{y}
    \right]
\end{aligned}
\end{equation}

where $ N_{x}, N_{y}, N_{z} $ correspond to the Cartesian components of the position of the nuclear center $N$ that the ket basis locates on,
\begin{equation}
    \boldsymbol{R}_{N} = (N_{x}, N_{y}, N_{z})
\end{equation}

and $ B_{x}, B_{y}, B_{z} $ correspond to the Cartesian components of external finite magnetic field strength,
\begin{equation}
    \boldsymbol{B} = (B_{x}, B_{y}, B_{z})
\end{equation}

\bibliography{refs}

\clearpage

\begingroup

\begin{minipage}{\textwidth}

{\large\bfseries Supporting Information for ``Efficient, precise DFT calculations of NMR shieldings: Revisiting the finite field approach'' \par}

\end{minipage}

\vspace{1.2em}

\setcounter{figure}{0}
\setcounter{table}{0}
\setcounter{equation}{0}
\setcounter{section}{0}

\renewcommand{\thefigure}{S\arabic{figure}}
\renewcommand{\thetable}{S\arabic{table}}
\renewcommand{\theequation}{S\arabic{equation}}
\renewcommand{\thesection}{S\arabic{section}}

\section{Error statistics for the numerical accuracy benchmark}

\begingroup
\scriptsize
\setlength{\tabcolsep}{6pt}
\renewcommand{\arraystretch}{1.0}

\setlength{\LTleft}{\fill}
\setlength{\LTright}{\fill}

\begin{longtable}[H]{
  l l l
  S[table-format=-1.4]
  S[table-format=1.3]
  S[table-format=1.3]
  S[table-format=1.3]
}

\caption{Errors in the Hartree-Fock shielding constant (unit: ppm) calculated from the finite difference first derivative (FD1) scheme versus the corresponding fully analytical results, for $^{7}$Li, $^{15}$N, $^{17}$O, $^{31}$P, and $^{33}$S}
\label{tab:si-fd1-errors} \\

\toprule
nuclear type & primary basis & computational settings
& {MSE} & {MAE} & {RMSE} & {MaxAE} \\
\midrule
\endfirsthead

\caption[]{Errors in the Hartree-Fock shielding constant (unit: ppm) calculated from the finite difference first derivative (FD1) scheme versus the corresponding fully analytical results, for $^{7}$Li, $^{15}$N, $^{17}$O, $^{31}$P, and $^{33}$S, continued} \\

\toprule
nuclear type & primary basis & computational settings
& {MSE} & {MAE} & {RMSE} & {MaxAE} \\
\midrule
\endhead

\endfoot

\multirow{10}{*}{$^{7}$Li (2 data points)}
& \multirow{7}{*}{pcSseg-2}
& exact, $1\times10^{-3}$ a.u. & 0.000 & 0.000 & 0.000 & 0.000 \\*
& & exact, $5\times10^{-4}$ a.u. & 0.000 & 0.000 & 0.000 & 0.000 \\*
& & exact, $1\times10^{-4}$ a.u. & 0.000 & 0.000 & 0.000 & 0.000 \\*
& & RI, $1\times10^{-3}$ a.u. & 0.000 & 0.000 & 0.000 & 0.000 \\*
& & RI, $5\times10^{-4}$ a.u. & 0.000 & 0.000 & 0.000 & 0.000 \\*
& & RI, $1\times10^{-4}$ a.u. & 0.000 & 0.000 & 0.000 & 0.000 \\*
\cline{2-7}

& \multirow{3}{*}{pcSseg-3}
& RI, $1\times10^{-3}$ a.u. & 0.000 & 0.000 & 0.000 & 0.000 \\*
& & RI, $5\times10^{-4}$ a.u. & 0.000 & 0.000 & 0.000 & 0.000 \\*
& & RI, $1\times10^{-4}$ a.u. & 0.000 & 0.000 & 0.000 & 0.000 \\*

\midrule

\multirow{10}{*}{$^{15}$N (7 data points)}
& \multirow{7}{*}{pcSseg-2}
& exact, $1\times10^{-3}$ a.u. & 0.002 & 0.002 & 0.004 & 0.010 \\*
& & exact, $5\times10^{-4}$ a.u. & 0.000 & 0.001 & 0.001 & 0.003 \\*
& & exact, $1\times10^{-4}$ a.u. & 0.000 & 0.000 & 0.001 & 0.001 \\*
& & RI, $1\times10^{-3}$ a.u. & 0.002 & 0.003 & 0.004 & 0.011 \\*
& & RI, $5\times10^{-4}$ a.u. & 0.001 & 0.001 & 0.001 & 0.003 \\*
& & RI, $1\times10^{-4}$ a.u. & 0.000 & 0.001 & 0.001 & 0.003 \\*
\cline{2-7}

& \multirow{3}{*}{pcSseg-3}
& RI, $1\times10^{-3}$ a.u. & 0.002 & 0.002 & 0.004 & 0.010 \\*
& & RI, $5\times10^{-4}$ a.u. & 0.001 & 0.001 & 0.001 & 0.003 \\*
& & RI, $1\times10^{-4}$ a.u. & 0.000 & 0.001 & 0.001 & 0.003 \\*

\midrule

\multirow{10}{*}{$^{17}$O (11 data points)}
& \multirow{7}{*}{pcSseg-2}
& exact, $1\times10^{-3}$ a.u. & 0.002 & 0.003 & 0.005 & 0.012 \\*
& & exact, $5\times10^{-4}$ a.u. & 0.000 & 0.001 & 0.001 & 0.003 \\*
& & exact, $1\times10^{-4}$ a.u. & -0.001 & 0.002 & 0.002 & 0.010 \\*
& & RI, $1\times10^{-3}$ a.u. & 0.002 & 0.003 & 0.005 & 0.012 \\*
& & RI, $5\times10^{-4}$ a.u. & 0.001 & 0.001 & 0.001 & 0.003 \\*
& & RI, $1\times10^{-4}$ a.u. & -0.002 & 0.003 & 0.004 & 0.010 \\*
\cline{2-7}

& \multirow{3}{*}{pcSseg-3}
& RI, $1\times10^{-3}$ a.u. & 0.002 & 0.003 & 0.005 & 0.012 \\*
& & RI, $5\times10^{-4}$ a.u. & 0.001 & 0.001 & 0.002 & 0.004 \\*
& & RI, $1\times10^{-4}$ a.u. & 0.001 & 0.005 & 0.008 & 0.023 \\*

\midrule

\multirow{10}{*}{$^{31}$P (2 data points)}
& \multirow{7}{*}{pcSseg-2}
& exact, $1\times10^{-3}$ a.u. & 0.010 & 0.010 & 0.012 & 0.016 \\*
& & exact, $5\times10^{-4}$ a.u. & 0.002 & 0.002 & 0.003 & 0.004 \\*
& & exact, $1\times10^{-4}$ a.u. & 0.001 & 0.001 & 0.001 & 0.001 \\*
& & RI, $1\times10^{-3}$ a.u. & 0.010 & 0.010 & 0.012 & 0.016 \\*
& & RI, $5\times10^{-4}$ a.u. & 0.002 & 0.002 & 0.003 & 0.004 \\*
& & RI, $1\times10^{-4}$ a.u. & 0.001 & 0.003 & 0.003 & 0.004 \\*
\cline{2-7}

& \multirow{3}{*}{pcSseg-3}
& RI, $1\times10^{-3}$ a.u. & 0.010 & 0.010 & 0.011 & 0.015 \\*
& & RI, $5\times10^{-4}$ a.u. & 0.002 & 0.002 & 0.002 & 0.003 \\*
& & RI, $1\times10^{-4}$ a.u. & 0.000 & 0.002 & 0.002 & 0.002 \\*

\midrule

\multirow{10}{*}{$^{33}$S (3 data points)}
& \multirow{7}{*}{pcSseg-2}
& exact, $1\times10^{-3}$ a.u. & 0.004 & 0.007 & 0.009 & 0.016 \\*
& & exact, $5\times10^{-4}$ a.u. & 0.002 & 0.002 & 0.003 & 0.005 \\*
& & exact, $1\times10^{-4}$ a.u. & 0.000 & 0.003 & 0.003 & 0.004 \\*
& & RI, $1\times10^{-3}$ a.u. & 0.003 & 0.006 & 0.008 & 0.014 \\*
& & RI, $5\times10^{-4}$ a.u. & 0.001 & 0.002 & 0.002 & 0.004 \\*
& & RI, $1\times10^{-4}$ a.u. & 0.003 & 0.003 & 0.004 & 0.006 \\*
\cline{2-7}

& \multirow{3}{*}{pcSseg-3}
& RI, $1\times10^{-3}$ a.u. & 0.002 & 0.006 & 0.008 & 0.012 \\*
& & RI, $5\times10^{-4}$ a.u. & 0.001 & 0.002 & 0.002 & 0.004 \\*
& & RI, $1\times10^{-4}$ a.u. & 0.004 & 0.004 & 0.005 & 0.008 \\

\end{longtable}
\endgroup

\begingroup
\scriptsize
\setlength{\tabcolsep}{6pt}
\renewcommand{\arraystretch}{1.0}

\setlength{\LTleft}{\fill}
\setlength{\LTright}{\fill}

\begin{longtable}[H]{
  l l l
  S[table-format=-1.4]
  S[table-format=1.3]
  S[table-format=1.3]
  S[table-format=1.3]
}

\caption{Overall combined errors in the Hartree-Fock shielding constant (unit: ppm) relative to the fully analytical results with exact JK build, for $^{7}$Li, $^{15}$N, $^{17}$O, $^{31}$P, and $^{33}$S}
\label{tab:overall-errors-hcf} \\

\toprule
nuclear type & primary basis & computational settings
& {MSE} & {MAE} & {RMSE} & {MaxAE} \\
\midrule
\endfirsthead

\caption[]{Overall combined errors in the Hartree-Fock shielding constant (unit: ppm) relative to the fully analytical results with exact JK build, for $^{7}$Li, $^{15}$N, $^{17}$O, $^{31}$P, and $^{33}$S, continued} \\

\toprule
nuclear type & primary basis & computational settings
& {MSE} & {MAE} & {RMSE} & {MaxAE} \\
\midrule
\endhead

\endfoot

\multirow{12}{*}{$^{7}$Li (2 data points)}
& \multirow{6}{*}{pcSseg-2}
& FD1, RI/def2-JK, $1\times10^{-3}$ a.u. & 0.001 & 0.001 & 0.001 & 0.001 \\*
& & FD1, RI/def2-JK, $5\times10^{-4}$ a.u. & 0.001 & 0.001 & 0.001 & 0.001 \\*
& & FD1, RI/def2-JK, $1\times10^{-4}$ a.u. & 0.001 & 0.001 & 0.001 & 0.001 \\*
& & analytical, RIJCOSX/def2-J & 0.012 & 0.012 & 0.012 & 0.012 \\*
& & analytical, RIJCOSX/def2-JK & 0.002 & 0.002 & 0.002 & 0.002 \\*
& & analytical, RI-JK/def2-JK & 0.001 & 0.001 & 0.001 & 0.001 \\*
\cline{2-7}

& \multirow{6}{*}{pcSseg-3}
& FD1, RI/def2-JK, $1\times10^{-3}$ a.u. & 0.002 & 0.002 & 0.002 & 0.002 \\*
& & FD1, RI/def2-JK, $5\times10^{-4}$ a.u. & 0.002 & 0.002 & 0.002 & 0.002 \\*
& & FD1, RI/def2-JK, $1\times10^{-4}$ a.u. & 0.002 & 0.002 & 0.002 & 0.002 \\*
& & analytical, RIJCOSX/def2-J & 0.013 & 0.013 & 0.013 & 0.014 \\*
& & analytical, RIJCOSX/def2-JK & 0.002 & 0.002 & 0.002 & 0.003 \\*
& & analytical, RI-JK/def2-JK & 0.002 & 0.002 & 0.002 & 0.002 \\*

\midrule

\multirow{12}{*}{$^{15}$N (7 data points)}
& \multirow{6}{*}{pcSseg-2}
& FD1, RI/def2-JK, $1\times10^{-3}$ a.u. & -0.006 & 0.009 & 0.013 & 0.032 \\*
& & FD1, RI/def2-JK, $5\times10^{-4}$ a.u. & -0.007 & 0.010 & 0.016 & 0.040 \\*
& & FD1, RI/def2-JK, $1\times10^{-4}$ a.u. & -0.008 & 0.010 & 0.017 & 0.042  \\*
& & analytical, RIJCOSX/def2-J & 0.127 & 0.151 & 0.320 & 0.838 \\*
& & analytical, RIJCOSX/def2-JK & -0.004 & 0.006 & 0.009 & 0.023 \\*
& & analytical, RI-JK/def2-JK & -0.008 & 0.011 & 0.017 & 0.042 \\*
\cline{2-7}

& \multirow{6}{*}{pcSseg-3}
& FD1, RI/def2-JK, $1\times10^{-3}$ a.u. & 0.006 & 0.015 & 0.017 & 0.027 \\*
& & FD1, RI/def2-JK, $5\times10^{-4}$ a.u. & 0.005 & 0.015 & 0.018 & 0.034 \\*
& & FD1, RI/def2-JK, $1\times10^{-4}$ a.u. & 0.004 & 0.016 & 0.019 & 0.037  \\*
& & analytical, RIJCOSX/def2-J & 0.156 & 0.164 & 0.361 & 0.948 \\*
& & analytical, RIJCOSX/def2-JK & 0.006 & 0.006 & 0.007 & 0.012 \\*
& & analytical, RI-JK/def2-JK & 0.005 & 0.016 & 0.018 & 0.037 \\*

\midrule

\multirow{12}{*}{$^{17}$O (11 data points)}
& \multirow{6}{*}{pcSseg-2}
& FD1, RI/def2-JK, $1\times10^{-3}$ a.u. & -0.041 & 0.043 & 0.074 & 0.229 \\*
& & FD1, RI/def2-JK, $5\times10^{-4}$ a.u. & -0.043 & 0.044 & 0.076 & 0.237 \\*
& & FD1, RI/def2-JK, $1\times10^{-4}$ a.u. & -0.045 & 0.046 & 0.078 & 0.240 \\*
& & analytical, RIJCOSX/def2-J & 0.237 & 0.265 & 0.458 & 1.126 \\*
& & analytical, RIJCOSX/def2-JK & 0.021 & 0.031 & 0.045 & 0.101 \\*
& & analytical, RI-JK/def2-JK & -0.043 & 0.045 & 0.077 & 0.240 \\*
\cline{2-7}

& \multirow{6}{*}{pcSseg-3}
& FD1, RI/def2-JK, $1\times10^{-3}$ a.u. & -0.031 & 0.043 & 0.073 & 0.229 \\*
& & FD1, RI/def2-JK, $5\times10^{-4}$ a.u. & -0.032 & 0.042 & 0.075 & 0.236 \\*
& & FD1, RI/def2-JK, $1\times10^{-4}$ a.u. & -0.032 & 0.041 & 0.076 & 0.241 \\*
& & analytical, RIJCOSX/def2-J & 0.215 & 0.237 & 0.415 & 0.929 \\*
& & analytical, RIJCOSX/def2-JK & 0.024 & 0.025 & 0.040 & 0.098 \\*
& & analytical, RI-JK/def2-JK & -0.033 & 0.042 & 0.076 & 0.238 \\*

\midrule

\multirow{12}{*}{$^{31}$P (2 data points)}
& \multirow{6}{*}{pcSseg-2}
& FD1, RI/def2-JK, $1\times10^{-3}$ a.u. & -0.058 & 0.058 & 0.060 & 0.075 \\*
& & FD1, RI/def2-JK, $5\times10^{-4}$ a.u. & -0.066 & 0.066 & 0.069 & 0.088 \\*
& & FD1, RI/def2-JK, $1\times10^{-4}$ a.u. & -0.067 & 0.067 & 0.070 & 0.088 \\*
& & analytical, RIJCOSX/def2-J & 0.507 & 0.507 & 0.641 & 0.899 \\*
& & analytical, RIJCOSX/def2-JK & 0.016 & 0.016 & 0.018 & 0.024 \\*
& & analytical, RI-JK/def2-JK & -0.068 & 0.068 & 0.072 & 0.091 \\*
\cline{2-7}

& \multirow{6}{*}{pcSseg-3}
& FD1, RI/def2-JK, $1\times10^{-3}$ a.u. & -0.084 & 0.084 & 0.089 & 0.114 \\*
& & FD1, RI/def2-JK, $5\times10^{-4}$ a.u. & -0.091 & 0.091 & 0.098 & 0.127 \\*
& & FD1, RI/def2-JK, $1\times10^{-4}$ a.u. & -0.093 & 0.093 & 0.099 & 0.127 \\*
& & analytical, RIJCOSX/def2-J & 0.609 & 0.609 & 0.737 & 1.024 \\*
& & analytical, RIJCOSX/def2-JK & 0.017 & 0.017 & 0.017 & 0.018 \\*
& & analytical, RI-JK/def2-JK & -0.093 & 0.093 & 0.099 & 0.129 \\*

\midrule

\multirow{12}{*}{$^{33}$S (3 data points)}
& \multirow{6}{*}{pcSseg-2}
& FD1, RI/def2-JK, $1\times10^{-3}$ a.u. & -0.154 & 0.154 & 0.191 & 0.311 \\*
& & FD1, RI/def2-JK, $5\times10^{-4}$ a.u. & -0.156 & 0.156 & 0.196 & 0.321 \\*
& & FD1, RI/def2-JK, $1\times10^{-4}$ a.u. & -0.155 & 0.155 & 0.196 & 0.324 \\*
& & analytical, RIJCOSX/def2-J & 0.226 & 0.372 & 0.531 & 0.896 \\*
& & analytical, RIJCOSX/def2-JK & 0.045 & 0.046 & 0.056 & 0.071 \\*
& & analytical, RI-JK/def2-JK & -0.157 & 0.157 & 0.198 & 0.325 \\*
\cline{2-7}

& \multirow{6}{*}{pcSseg-3}
& FD1, RI/def2-JK, $1\times10^{-3}$ a.u. & -0.192 & 0.192 & 0.249 & 0.412 \\*
& & FD1, RI/def2-JK, $5\times10^{-4}$ a.u. & -0.193 & 0.193 & 0.253 & 0.420 \\*
& & FD1, RI/def2-JK, $1\times10^{-4}$ a.u. & -0.191 & 0.191 & 0.253 & 0.421 \\*
& & analytical, RIJCOSX/def2-J & 0.196 & 0.312 & 0.444 & 0.749 \\*
& & analytical, RIJCOSX/def2-JK & -0.001 & 0.007 & 0.007 & 0.008 \\*
& & analytical, RI-JK/def2-JK &  -0.194 & 0.194 & 0.255 & 0.424 \\*

\end{longtable}
\endgroup

\begingroup
\scriptsize
\setlength{\tabcolsep}{3pt}
\renewcommand{\arraystretch}{1.0}

\setlength{\LTleft}{\fill}
\setlength{\LTright}{\fill}

\begin{longtable}[H]{
  l l l
  S[table-format=-1.4]
  S[table-format=1.3]
  S[table-format=1.3]
  S[table-format=1.3]
}

\caption{Overall combined errors in the TPSSh shielding constant (unit: ppm) relative to the fully analytical results with exact JK build, for all elements}
\label{tab:overall-errors} \\

\toprule
nuclear type & primary basis & computational settings
& {MSE} & {MAE} & {RMSE} & {MaxAE} \\
\midrule
\endfirsthead

\caption[]{Overall combined errors in the TPSSh shielding constant (unit: ppm) relative to the fully analytical results with exact JK build, for all elements, continued} \\

\toprule
nuclear type & primary basis & computational settings
& {MSE} & {MAE} & {RMSE} & {MaxAE} \\
\midrule
\endhead

\endfoot

\multirow{18}{*}{$^{1}$H (18 data points)}
& \multirow{9}{*}{pcSseg-2}
& FD1, RI/def2-JK, default SG-2 grid, $1*10^{-3}$ a.u. & 0.003 & 0.004 & 0.006 & 0.016 \\*
& & FD1, RI/def2-JK, default SG-2 grid, $5*10^{-4}$ a.u. & 0.003 & 0.004 & 0.006 & 0.016 \\*
& & FD1, RI/def2-JK, default SG-2 grid, $1*10^{-4}$ a.u. & 0.003 & 0.004 & 0.006 & 0.016 \\*
& & FD1, RI/def2-JK, unpruned (75,302) grid, $1*10^{-3}$ a.u. & 0.001 & 0.001 & 0.001 & 0.002 \\*
& & FD1, RI/def2-JK, unpruned (75,302) grid, $5*10^{-4}$ a.u. & 0.001 & 0.001 & 0.001 & 0.002 \\*
& & FD1, RI/def2-JK, unpruned (75,302) grid, $1*10^{-4}$ a.u. & 0.001 & 0.001 & 0.001 & 0.002 \\*
& & analytical, RIJCOSX/def2-J & -0.002 & 0.003 & 0.004 & 0.013 \\*
& & analytical, RIJCOSX/def2-JK & 0.001 & 0.001 & 0.002 & 0.002 \\*
& & analytical, RI-JK/def2-JK & 0.001 & 0.001 & 0.001 & 0.002 \\*
\cline{2-7}

& \multirow{9}{*}{pcSseg-3}
& FD1, RI/def2-JK, default SG-2 grid, $1*10^{-3}$ a.u. & 0.003 & 0.004 & 0.006 & 0.015 \\*
& & FD1, RI/def2-JK, default SG-2 grid, $5*10^{-4}$ a.u. & 0.003 & 0.004 & 0.006 & 0.015 \\*
& & FD1, RI/def2-JK, default SG-2 grid, $1*10^{-4}$ a.u. & 0.003 & 0.004 & 0.006 & 0.015 \\*
& & FD1, RI/def2-JK, unpruned (75,302) grid, $1*10^{-3}$ a.u. & 0.001 & 0.001 & 0.001 & 0.002 \\*
& & FD1, RI/def2-JK, unpruned (75,302) grid, $5*10^{-4}$ a.u. & 0.001 & 0.001 & 0.001 & 0.002 \\*
& & FD1, RI/def2-JK, unpruned (75,302) grid, $1*10^{-4}$ a.u. & 0.001 & 0.001 & 0.001 & 0.002 \\*
& & analytical, RIJCOSX/def2-J & -0.002 & 0.003 & 0.004 & 0.012 \\*
& & analytical, RIJCOSX/def2-JK & 0.001 & 0.001 & 0.002 & 0.002 \\*
& & analytical, RI-JK/def2-JK & 0.001 & 0.001 & 0.001 & 0.002 \\*

\midrule

\multirow{18}{*}{$^{7}$Li (2 data points)}
& \multirow{9}{*}{pcSseg-2}
& FD1, RI/def2-JK, default SG-2 grid, $1*10^{-3}$ a.u. & 0.001 & 0.001 & 0.001 & 0.002 \\*
& & FD1, RI/def2-JK, default SG-2 grid, $5*10^{-4}$ a.u. & 0.001 & 0.001 & 0.001 & 0.002 \\*
& & FD1, RI/def2-JK, default SG-2 grid, $1*10^{-4}$ a.u. & 0.001 & 0.001 & 0.001 & 0.002 \\*
& & FD1, RI/def2-JK, unpruned (75,302) grid, $1*10^{-3}$ a.u. & 0.002 & 0.002 & 0.003 & 0.003 \\*
& & FD1, RI/def2-JK, unpruned (75,302) grid, $5*10^{-4}$ a.u. & 0.003 & 0.003 & 0.003 & 0.003 \\*
& & FD1, RI/def2-JK, unpruned (75,302) grid, $1*10^{-4}$ a.u. & 0.003 & 0.003 & 0.003 & 0.003 \\*
& & analytical, RIJCOSX/def2-J & 0.016 & 0.016 & 0.016 & 0.016 \\*
& & analytical, RIJCOSX/def2-JK & 0.003 & 0.003 & 0.003 & 0.003 \\*
& & analytical, RI-JK/def2-JK & 0.001 & 0.001 & 0.001 & 0.001 \\*
\cline{2-7}

& \multirow{9}{*}{pcSseg-3}
& FD1, RI/def2-JK, default SG-2 grid, $1*10^{-3}$ a.u. & 0.002 & 0.002 & 0.003 & 0.004 \\*
& & FD1, RI/def2-JK, default SG-2 grid, $5*10^{-4}$ a.u. & 0.002 & 0.002 & 0.003 & 0.004 \\*
& & FD1, RI/def2-JK, default SG-2 grid, $1*10^{-4}$ a.u. & 0.002 & 0.002 & 0.003 & 0.004 \\*
& & FD1, RI/def2-JK, unpruned (75,302) grid, $1*10^{-3}$ a.u. & 0.003 & 0.003 & 0.003 & 0.004 \\*
& & FD1, RI/def2-JK, unpruned (75,302) grid, $5*10^{-4}$ a.u. & 0.003 & 0.003 & 0.003 & 0.004 \\*
& & FD1, RI/def2-JK, unpruned (75,302) grid, $1*10^{-4}$ a.u. & 0.003 & 0.003 & 0.003 & 0.004 \\*
& & analytical, RIJCOSX/def2-J & 0.017 & 0.017 & 0.017 & 0.018 \\*
& & analytical, RIJCOSX/def2-JK & 0.003 & 0.003 & 0.003 & 0.004 \\*
& & analytical, RI-JK/def2-JK & 0.003 & 0.003 & 0.003 & 0.004 \\*

\midrule

\multirow{18}{*}{$^{13}$C (17 data points)}
& \multirow{9}{*}{pcSseg-2}
& FD1, RI/def2-JK, default SG-2 grid, $1*10^{-3}$ a.u. & 0.015 & 0.023 & 0.036 & 0.122 \\*
& & FD1, RI/def2-JK, default SG-2 grid, $5*10^{-4}$ a.u. & 0.015 & 0.023 & 0.035 & 0.121 \\*
& & FD1, RI/def2-JK, default SG-2 grid, $1*10^{-4}$ a.u. & 0.015 & 0.023 & 0.035 & 0.120 \\*
& & FD1, RI/def2-JK, unpruned (75,302) grid, $1*10^{-3}$ a.u. & 0.003 & 0.005 & 0.007 & 0.017 \\*
& & FD1, RI/def2-JK, unpruned (75,302) grid, $5*10^{-4}$ a.u. & 0.003 & 0.005 & 0.007 & 0.017 \\*
& & FD1, RI/def2-JK, unpruned (75,302) grid, $1*10^{-4}$ a.u. & 0.003 & 0.005 & 0.007 & 0.018 \\*
& & analytical, RIJCOSX/def2-J & 0.010 & 0.013 & 0.022 & 0.069 \\*
& & analytical, RIJCOSX/def2-JK & 0.004 & 0.006 & 0.008 & 0.020 \\*
& & analytical, RI-JK/def2-JK & 0.004 & 0.005 & 0.007 & 0.016 \\*
\cline{2-7}

& \multirow{9}{*}{pcSseg-3}
& FD1, RI/def2-JK, default SG-2 grid, $1*10^{-3}$ a.u. & 0.011 & 0.022 & 0.033 & 0.112 \\*
& & FD1, RI/def2-JK, default SG-2 grid, $5*10^{-4}$ a.u. & 0.011 & 0.022 & 0.033 & 0.111 \\*
& & FD1, RI/def2-JK, default SG-2 grid, $1*10^{-4}$ a.u. & 0.011 & 0.022 & 0.032 & 0.109 \\*
& & FD1, RI/def2-JK, unpruned (75,302) grid, $1*10^{-3}$ a.u. & 0.003 & 0.003 & 0.003 & 0.004 \\*
& & FD1, RI/def2-JK, unpruned (75,302) grid, $5*10^{-4}$ a.u. & 0.003 & 0.003 & 0.003 & 0.004 \\*
& & FD1, RI/def2-JK, unpruned (75,302) grid, $1*10^{-4}$ a.u. & 0.003 & 0.003 & 0.003 & 0.004 \\*
& & analytical, RIJCOSX/def2-J & 0.017 & 0.017 & 0.017 & 0.018 \\*
& & analytical, RIJCOSX/def2-JK & 0.003 & 0.003 & 0.003 & 0.004 \\*
& & analytical, RI-JK/def2-JK & 0.003 & 0.003 & 0.003 & 0.004 \\*

\midrule

\multirow{18}{*}{$^{15}$N (7 data points)}
& \multirow{9}{*}{pcSseg-2}
& FD1, RI/def2-JK, default SG-2 grid, $1*10^{-3}$ a.u. & 0.078 & 0.086 & 0.155 & 0.367 \\*
& & FD1, RI/def2-JK, default SG-2 grid, $5*10^{-4}$ a.u. & 0.076 & 0.084 & 0.152 & 0.357 \\*
& & FD1, RI/def2-JK, default SG-2 grid, $1*10^{-4}$ a.u. & 0.075 & 0.053 & 0.150 & 0.354 \\*
& & FD1, RI/def2-JK, unpruned (75,302) grid, $1*10^{-3}$ a.u. & -0.004 & 0.007 & 0.012 & 0.029 \\*
& & FD1, RI/def2-JK, unpruned (75,302) grid, $5*10^{-4}$ a.u. & -0.006 & 0.009 & 0.015 & 0.039 \\*
& & FD1, RI/def2-JK, unpruned (75,302) grid, $1*10^{-4}$ a.u. & -0.007 & 0.010 & 0.017 & 0.042 \\*
& & analytical, RIJCOSX/def2-J & 0.091 & 0.111 & 0.222 & 0.579 \\*
& & analytical, RIJCOSX/def2-JK & -0.004 & 0.006 & 0.008 & 0.019 \\*
& & analytical, RI-JK/def2-JK & 0.000 & 0.003 & 0.004 & 0.009 \\*
\cline{2-7}

& \multirow{9}{*}{pcSseg-3}
& FD1, RI/def2-JK, default SG-2 grid, $1*10^{-3}$ a.u. & 0.066 & 0.076 & 0.131 & 0.293 \\*
& & FD1, RI/def2-JK, default SG-2 grid, $5*10^{-4}$ a.u. & 0.064 & 0.074 & 0.121 & 0.283 \\*
& & FD1, RI/def2-JK, default SG-2 grid, $1*10^{-4}$ a.u. & 0.063 & 0.074 & 0.126 & 0.280 \\*
& & FD1, RI/def2-JK, unpruned (75,302) grid, $1*10^{-3}$ a.u. & 0.004 & 0.006 & 0.006 & 0.010 \\*
& & FD1, RI/def2-JK, unpruned (75,302) grid, $5*10^{-4}$ a.u. & 0.002 & 0.007 & 0.008 & 0.016 \\*
& & FD1, RI/def2-JK, unpruned (75,302) grid, $1*10^{-4}$ a.u. & 0.002 & 0.007 & 0.009 & 0.019 \\*
& & analytical, RIJCOSX/def2-J & 0.121 & 0.131 & 0.276 & 0.723 \\*
& & analytical, RIJCOSX/def2-JK & 0.003 & 0.005 & 0.006 & 0.011 \\*
& & analytical, RI-JK/def2-JK & 0.007 & 0.007 & 0.007 & 0.011 \\*

\midrule

\multirow{18}{*}{$^{17}$O (11 data points)}
& \multirow{9}{*}{pcSseg-2}
& FD1, RI/def2-JK, default SG-2 grid, $1*10^{-3}$ a.u. & 0.036 & 0.069 & 0.091 & 0.184 \\*
& & FD1, RI/def2-JK, default SG-2 grid, $5*10^{-4}$ a.u. & 0.034 & 0.069 & 0.092 & 0.185 \\*
& & FD1, RI/def2-JK, default SG-2 grid, $1*10^{-4}$ a.u. & 0.034 & 0.069 & 0.092 & 0.186 \\*
& & FD1, RI/def2-JK, unpruned (75,302) grid, $1*10^{-3}$ a.u. & 0.002 & 0.012 & 0.018 & 0.050 \\*
& & FD1, RI/def2-JK, unpruned (75,302) grid, $5*10^{-4}$ a.u. & 0.001 & 0.014 & 0.019 & 0.051 \\*
& & FD1, RI/def2-JK, unpruned (75,302) grid, $1*10^{-4}$ a.u. & 0.000 & 0.014 & 0.019 & 0.051 \\*
& & analytical, RIJCOSX/def2-J & 0.135 & 0.161 & 0.293 & 0.800 \\*
& & analytical, RIJCOSX/def2-JK & 0.005 & 0.015 & 0.024 & 0.070 \\*
& & analytical, RI-JK/def2-JK & 0.005 & 0.009 & 0.015 & 0.044 \\*
\cline{2-7}

& \multirow{9}{*}{pcSseg-3}
& FD1, RI/def2-JK, default SG-2 grid, $1*10^{-3}$ a.u. & 0.025 & 0.061 & 0.084 & 0.167 \\*
& & FD1, RI/def2-JK, default SG-2 grid, $5*10^{-4}$ a.u. & 0.023 & 0.062 & 0.085 & 0.176 \\*
& & FD1, RI/def2-JK, default SG-2 grid, $1*10^{-4}$ a.u. & 0.023 & 0.062 & 0.085 & 0.179 \\*
& & FD1, RI/def2-JK, unpruned (75,302) grid, $1*10^{-3}$ a.u. & 0.017 & 0.017 & 0.031 & 0.093 \\*
& & FD1, RI/def2-JK, unpruned (75,302) grid, $5*10^{-4}$ a.u. & 0.015 & 0.017 & 0.031 & 0.094 \\*
& & FD1, RI/def2-JK, unpruned (75,302) grid, $1*10^{-4}$ a.u. & 0.015 & 0.017 & 0.032 & 0.095 \\*
& & analytical, RIJCOSX/def2-J & 0.127 & 0.156 & 0.276 & 0.780 \\*
& & analytical, RIJCOSX/def2-JK & 0.018 & 0.021 & 0.037 & 0.109 \\*
& & analytical, RI-JK/def2-JK & 0.019 & 0.019 & 0.029 & 0.082 \\*

\midrule

\multirow{18}{*}{$^{19}$F (9 data points)}
& \multirow{9}{*}{pcSseg-2}
& FD1, RI/def2-JK, default SG-2 grid, $1*10^{-3}$ a.u. & 0.034 & 0.059 & 0.078 & 0.142 \\*
& & FD1, RI/def2-JK, default SG-2 grid, $5*10^{-4}$ a.u. & 0.034 & 0.059 & 0.077 & 0.143 \\*
& & FD1, RI/def2-JK, default SG-2 grid, $1*10^{-4}$ a.u. & 0.034 & 0.059 & 0.077 & 0.143 \\*
& & FD1, RI/def2-JK, unpruned (75,302) grid, $1*10^{-3}$ a.u. & -0.016 & 0.022 & 0.027 & 0.063 \\*
& & FD1, RI/def2-JK, unpruned (75,302) grid, $5*10^{-4}$ a.u. & -0.016 & 0.021 & 0.027 & 0.062 \\*
& & FD1, RI/def2-JK, unpruned (75,302) grid, $1*10^{-4}$ a.u. & -0.017 & 0.021 & 0.027 & 0.062 \\*
& & analytical, RIJCOSX/def2-J & 0.030 & 0.088 & 0.132 & 0.298 \\*
& & analytical, RIJCOSX/def2-JK & -0.011 & 0.015 & 0.018 & 0.032 \\*
& & analytical, RI-JK/def2-JK & -0.007 & 0.010 & 0.013 & 0.026 \\*
\cline{2-7}

& \multirow{9}{*}{pcSseg-3}
& FD1, RI/def2-JK, default SG-2 grid, $1*10^{-3}$ a.u. & 0.025 & 0.058 & 0.073 & 0.155 \\*
& & FD1, RI/def2-JK, default SG-2 grid, $5*10^{-4}$ a.u. & 0.025 & 0.058 & 0.072 & 0.151 \\*
& & FD1, RI/def2-JK, default SG-2 grid, $1*10^{-4}$ a.u. & 0.025 & 0.057 & 0.072 & 0.150 \\*
& & FD1, RI/def2-JK, unpruned (75,302) grid, $1*10^{-3}$ a.u. & 0.006 & 0.015 & 0.019 & 0.045 \\*
& & FD1, RI/def2-JK, unpruned (75,302) grid, $5*10^{-4}$ a.u. & 0.006 & 0.014 & 0.018 & 0.042 \\*
& & FD1, RI/def2-JK, unpruned (75,302) grid, $1*10^{-4}$ a.u. & 0.006 & 0.014 & 0.018 & 0.041 \\*
& & analytical, RIJCOSX/def2-J & 0.036 & 0.085 & 0.125 & 0.262 \\*
& & analytical, RIJCOSX/def2-JK & 0.004 & 0.010 & 0.013 & 0.023 \\*
& & analytical, RI-JK/def2-JK & 0.005 & 0.010 & 0.012 & 0.024 \\*

\midrule

\multirow{18}{*}{$^{31}$P (2 data points)}
& \multirow{9}{*}{pcSseg-2}
& FD1, RI/def2-JK, default SG-2 grid, $1*10^{-3}$ a.u. & 0.125 & 0.400 & 0.419 & 0.525 \\*
& & FD1, RI/def2-JK, default SG-2 grid, $5*10^{-4}$ a.u. & 0.118 & 0.396 & 0.413 & 0.514 \\*
& & FD1, RI/def2-JK, default SG-2 grid, $1*10^{-4}$ a.u. & 0.116 & 0.395 & 0.411 & 0.511 \\*
& & FD1, RI/def2-JK, unpruned (75,302) grid, $1*10^{-3}$ a.u. & -0.007 & 0.025 & 0.026 & 0.032 \\*
& & FD1, RI/def2-JK, unpruned (75,302) grid, $5*10^{-4}$ a.u. & -0.015 & 0.029 & 0.032 & 0.043 \\*
& & FD1, RI/def2-JK, unpruned (75,302) grid, $1*10^{-4}$ a.u. & -0.017 & 0.030 & 0.034 & 0.047 \\*
& & analytical, RIJCOSX/def2-J & 0.392 & 0.392 & 0.493 & 0.690 \\*
& & analytical, RIJCOSX/def2-JK & -0.009 & 0.014 & 0.017 & 0.023 \\*
& & analytical, RI-JK/def2-JK & 0.003 & 0.007 & 0.007 & 0.010 \\*
\cline{2-7}

& \multirow{9}{*}{pcSseg-3}
& FD1, RI/def2-JK, default SG-2 grid, $1*10^{-3}$ a.u. & 0.030 & 0.365 & 0.366 & 0.394 \\*
& & FD1, RI/def2-JK, default SG-2 grid, $5*10^{-4}$ a.u. & 0.023 & 0.361 & 0.361 & 0.383 \\*
& & FD1, RI/def2-JK, default SG-2 grid, $1*10^{-4}$ a.u. & 0.022 & 0.358 & 0.359 & 0.380 \\*
& & FD1, RI/def2-JK, unpruned (75,302) grid, $1*10^{-3}$ a.u. & -0.009 & 0.021 & 0.023 & 0.041 \\*
& & FD1, RI/def2-JK, unpruned (75,302) grid, $5*10^{-4}$ a.u. & -0.016 & 0.025 & 0.029 & 0.041 \\*
& & FD1, RI/def2-JK, unpruned (75,302) grid, $1*10^{-4}$ a.u. & -0.016 & 0.028 & 0.032 & 0.044 \\*
& & analytical, RIJCOSX/def2-J & 0.471 & 0.471 & 0.584 & 0.816 \\*
& & analytical, RIJCOSX/def2-JK & -0.005 & 0.012 & 0.012 & 0.016 \\*
& & analytical, RI-JK/def2-JK & 0.009 & 0.009 & 0.009 & 0.011 \\*

\midrule

\multirow{18}{*}{$^{33}$S (3 data points)}
& \multirow{9}{*}{pcSseg-2}
& FD1, RI/def2-JK, default SG-2 grid, $1*10^{-3}$ a.u. & 0.036 & 0.036 & 0.045 & 0.068 \\*
& & FD1, RI/def2-JK, default SG-2 grid, $5*10^{-4}$ a.u. & 0.035 & 0.035 & 0.045 & 0.071 \\*
& & FD1, RI/def2-JK, default SG-2 grid, $1*10^{-4}$ a.u. & 0.035 & 0.035 & 0.045 & 0.072 \\*
& & FD1, RI/def2-JK, unpruned (75,302) grid, $1*10^{-3}$ a.u. & 0.013 & 0.020 & 0.021 & 0.030 \\*
& & FD1, RI/def2-JK, unpruned (75,302) grid, $5*10^{-4}$ a.u. & 0.012 & 0.018 & 0.021 & 0.033 \\*
& & FD1, RI/def2-JK, unpruned (75,302) grid, $1*10^{-4}$ a.u. & 0.011 & 0.017 & 0.021 & 0.034 \\*
& & analytical, RIJCOSX/def2-J & -0.033 & 0.166 & 0.203 & 0.288 \\*
& & analytical, RIJCOSX/def2-JK & 0.017 & 0.019 & 0.024 & 0.036 \\*
& & analytical, RI-JK/def2-JK & 0.005 & 0.010 & 0.013 & 0.021 \\*
\cline{2-7}

& \multirow{9}{*}{pcSseg-3}
& FD1, RI/def2-JK, default SG-2 grid, $1*10^{-3}$ a.u. & -0.002 & 0.085 & 0.095 & 0.130 \\*
& & FD1, RI/def2-JK, default SG-2 grid, $5*10^{-4}$ a.u. & -0.003 & 0.083 & 0.095 & 0.129 \\*
& & FD1, RI/def2-JK, default SG-2 grid, $1*10^{-4}$ a.u. & -0.002 & 0.082 & 0.094 & 0.126 \\*
& & FD1, RI/def2-JK, unpruned (75,302) grid, $1*10^{-3}$ a.u. & 0.015 & 0.023 & 0.029 & 0.047 \\*
& & FD1, RI/def2-JK, unpruned (75,302) grid, $5*10^{-4}$ a.u. & 0.015 & 0.022 & 0.030 & 0.051 \\*
& & FD1, RI/def2-JK, unpruned (75,302) grid, $1*10^{-4}$ a.u. & 0.015 & 0.020 & 0.030 & 0.052 \\*
& & analytical, RIJCOSX/def2-J & 0.026 & 0.129 & 0.158 & 0.226 \\*
& & analytical, RIJCOSX/def2-JK & 0.018 & 0.019 & 0.027 & 0.045 \\*
& & analytical, RI-JK/def2-JK & 0.010 & 0.016 & 0.022 & 0.038 \\*

\end{longtable}
\endgroup

\section{Single-node timing}

\begin{table}[H]
\hspace*{-0.75cm}
\begin{tabular}{@{}cccc@{}}
  & \shortstack{Benzene \\ \textbf{$N=630$}}   & \shortstack{Caffeine \\ \textbf{$N=1338$}} & \shortstack{Aspartame \\ \textbf{$N=2106$}}  \\ 
\midrule
\shortstack{ \\ MPI x OMP \\ 1 x 1} & \shortstack{FD1: 755 \\ Analytical RIJCOSX: 832 \\ Analytical RIJK: 366 } & \shortstack{FD1: 5206 \\ Analytical RIJCOSX: 5538 \\ Analytical RIJK: 3423 } & \shortstack{FD1: 17328 \\ Analytical RIJCOSX: 14869 \\ Analytical RIJK: 13944 } \\
\midrule
\shortstack{ \\ MPI x OMP \\ 1 x 4} & \shortstack{FD1: 208 \\ Analytical RIJCOSX: 266 \\ Analytical RIJK: 525 } & \shortstack{FD1: 1435 \\ Analytical RIJCOSX: 1601 \\ Analytical RIJK: 5935 } & \shortstack{FD1: 4683 \\ Analytical RIJCOSX: 3997 \\ Analytical RIJK: 10548 } \\
\midrule
\shortstack{ \\ MPI x OMP \\ 1 x 8} & \shortstack{FD1: 115 \\ Analytical RIJCOSX: 154 \\ Analytical RIJK: 385 } & \shortstack{FD1: 775 \\ Analytical RIJCOSX: 854 \\ Analytical RIJK: 2919 } & \shortstack{FD1: 2514 \\ Analytical RIJCOSX: 2087 \\ Analytical RIJK: 10197 } \\
\midrule
\shortstack{ \\ MPI x OMP \\ 1 x 16} & \shortstack{FD1: 72 \\ Analytical RIJCOSX: 117 \\ Analytical RIJK: 475 } & \shortstack{FD1: 478 \\ Analytical RIJCOSX: 565 \\ Analytical RIJK: 5394 } & \shortstack{FD1: 1576 \\ Analytical RIJCOSX: 1245 \\ Analytical RIJK: 10171 } \\
\midrule
\shortstack{ \\ MPI x OMP \\ 1 x 32} & \shortstack{FD1: 54 \\ Analytical RIJCOSX: 113 \\ Analytical RIJK: 491 } & \shortstack{FD1: 371 \\ Analytical RIJCOSX: 454 \\ Analytical RIJK: 10014 } & \shortstack{FD1: 1125 \\ Analytical RIJCOSX: 927 \\ Analytical RIJK: n/a} \\
\caption {Wall time (in seconds) comparison between analytical RIJCOSX (with def2-J), analytical RI-JK (with def2-JK) and FD1 (RI-J + occ-RI-K with def2-JK)  algorithms for chemical shielding using the specified computing resources (MPI ranks times OMP threads) for 3 larger systems, ATP, cholesterol, and lipitor, ran on a single physical node} 
\end{tabular}
\end{table}

\begin{table}[H]
\hspace*{-1cm}
\begin{tabular}{@{}cccc@{}}
  & \shortstack{ATP \\ \textbf{$N=2772$}}   & \shortstack{Cholesterol \\ \textbf{$N=3534$}} & \shortstack{Lipitor \\ \textbf{$N=4107$}}  \\ 
\midrule
\shortstack{MPI x OMP \\ 1 x 1} & \shortstack{FD1: 43665 \\ Analytical RIJCOSX: 27217} & \shortstack{FD1: 74634 \\ Analytical RIJCOSX: 43950} & \shortstack{FD1: 130567 \\ Analytical RIJCOSX: 69674} \\
\midrule
\shortstack{MPI x OMP \\ 1 x 4} & \shortstack{FD1: 11337 \\ Analytical RIJCOSX: 7155} & \shortstack{FD1: 20821 \\ Analytical RIJCOSX: n/a} & \shortstack{FD1: 36667 \\ Analytical RIJCOSX: n/a} \\
\midrule
\shortstack{MPI x OMP \\ 1 x 8} & \shortstack{FD1: 5883 \\ Analytical RIJCOSX: 3739} & \shortstack{FD1: 10471 \\ Analytical RIJCOSX: 6098} & \shortstack{FD1: 18213 \\ Analytical RIJCOSX: 9554} \\
\midrule
\shortstack{MPI x OMP \\ 1 x 16} & \shortstack{FD1: 3506 \\ Analytical RIJCOSX: 2170} & \shortstack{FD1: 6279 \\ Analytical RIJCOSX: 3457} & \shortstack{FD1: 11034 \\ Analytical RIJCOSX: 5412} \\
\midrule
\shortstack{MPI x OMP \\ 1 x 24} & \shortstack{FD1: 2887 \\ Analytical RIJCOSX: 1793} & \shortstack{FD1: 5087 \\ Analytical RIJCOSX: 2976} & \shortstack{FD1: 9426 \\ Analytical RIJCOSX: 4404} \\
\midrule
\shortstack{MPI x OMP \\ 1 x 32} & \shortstack{FD1: 2490 \\ Analytical RIJCOSX: 1542} & \shortstack{FD1: 4437 \\ Analytical RIJCOSX: 2445} & \shortstack{FD1: 8451 \\ Analytical RIJCOSX: 3741} \\
\midrule
\shortstack{MPI x OMP \\ 1 x 48} & \shortstack{FD1: n/a \\ Analytical RIJCOSX: n/a} & \shortstack{FD1: 3914 \\ Analytical RIJCOSX: 2220} & \shortstack{FD1: 7758 \\ Analytical RIJCOSX: 3307} \\
\midrule
\shortstack{MPI x OMP \\ 1 x 64} & \shortstack{FD1: n/a \\ Analytical RIJCOSX: n/a} & \shortstack{FD1: 3530 \\ Analytical RIJCOSX: 2295} & \shortstack{FD1: 7457 \\ Analytical RIJCOSX: 3178} \\
\caption {Wall time (in seconds) comparison between analytical (RIJCOSX/def2-J) and FD1 (RI-J + occ-RI-K with def2-JK) algorithms for chemical shielding using the specified computing resources (MPI ranks times OMP threads) for 3 larger systems, ATP, cholesterol, and lipitor, ran on a single physical node} 
\end{tabular}
\end{table}

\section{Multi-node timing}

\begin{table}[H]
\hspace*{-1cm}
\begin{tabular}{@{}cccc@{}}
  & \shortstack{ATP \\ \textbf{$N=2772$}}   & \shortstack{Cholesterol \\ \textbf{$N=3534$}} & \shortstack{Lipitor \\ \textbf{$N=4107$}}  \\ 
\midrule
\shortstack{MPI x OMP \\ 3 x 4} & \shortstack{FD1: 4520 \\ Analytical RIJCOSX: 2714} & \shortstack{FD1: 7219 \\ Analytical RIJCOSX: 4414} & \shortstack{FD1: 13320 \\ Analytical RIJCOSX: 6913} \\
\midrule
\shortstack{MPI x OMP \\ 3 x 8} & \shortstack{FD1: 2533 \\ Analytical RIJCOSX: 1692} & \shortstack{FD1: 3617 \\ Analytical RIJCOSX: 2677} & \shortstack{FD1: 6572 \\ Analytical RIJCOSX: 4106} \\
\midrule
\shortstack{MPI x OMP \\ 3 x 16} & \shortstack{FD1: 1508 \\ Analytical RIJCOSX: 1261} & \shortstack{FD1: 2140 \\ Analytical RIJCOSX: 2059} & \shortstack{FD1: 3911 \\ Analytical RIJCOSX: 3103} \\
\midrule
\shortstack{MPI x OMP \\ 3 x 32} & \shortstack{FD1: 1096 \\ Analytical RIJCOSX: 1463} & \shortstack{FD1: 1548 \\ Analytical RIJCOSX: 1751} & \shortstack{FD1: 3114 \\ Analytical RIJCOSX: 2934} \\
\caption {Wall time (in seconds) comparison between analytical and FD1 (RI-J + occ-RI-K with def2-JK) algorithms for chemical shielding using the specified computing resources (MPI ranks times OMP threads) for 3 larger systems, ATP, cholesterol, and lipitor, ran on 3 physical nodes} 
\end{tabular}
\end{table}

\endgroup

\end{document}